\documentclass[a4paper, 11pt]{article} 

\usepackage[protrusion=true,expansion=true]{microtype} 
\usepackage{graphicx} 
\usepackage{wrapfig} 
\usepackage{hyperref}

\usepackage{mathpazo} 
\usepackage[T1]{fontenc} 
\makeatletter
\renewcommand\@biblabel[1]{\textbf{#1.}} 
\renewcommand{\@listI}{\itemsep=0pt} 

\renewcommand{\maketitle}{ 
	\begin{center} 
		{\LARGE\@title} 
		
		\vspace{50pt} 
		
		{\large\@author} 
		\\\vspace{5pt}\@date 
		
		\vspace{40pt} 
	\end{center}
}

\usepackage{graphicx}
\usepackage{amsmath}
\usepackage{mathtools}
\usepackage{dsfont}
\usepackage{color}
\usepackage[mathscr]{euscript}
\usepackage{comment}
\usepackage{graphicx}
\usepackage{placeins}
\usepackage{float}
\usepackage{multirow}
\usepackage{amssymb}
\usepackage{pdflscape}
\usepackage{caption}
\usepackage{subcaption}
\usepackage{tikz}
\usepackage{lineno}
\usetikzlibrary{decorations.pathreplacing}
\usetikzlibrary{backgrounds} 
\usetikzlibrary{calc}
\pgfdeclarelayer{bg}    
\pgfsetlayers{bg,main}
\usepackage{rotate}
\usepackage{tikz}
\usetikzlibrary{decorations.pathreplacing}
\usetikzlibrary{calligraphy}
\usetikzlibrary{backgrounds} 
\usetikzlibrary{calc}

\usetikzlibrary{arrows,calc}
\tikzset{
	>=stealth',
	help lines/.style={dashed, thick},
	axis/.style={<->},
	important line/.style={thick},
	connection/.style={thick, dotted},
}

\allowdisplaybreaks
\newtheorem{theorem}{Theorem}[section]
\newtheorem{remark}[theorem]{Remark}
\newenvironment{proof}{\paragraph{Proof of Theorem \ref{main_theorem}:}}{\hfill$\blacksquare$}

\newtheorem{definition}[theorem]{Definition}

\tikzset{
	square/.style={%
		draw=none,
		circle,
		append after command={%
			\pgfextra \draw[black] (\tikzlastnode.north-|\tikzlastnode.west) rectangle 
			(\tikzlastnode.south-|\tikzlastnode.east);\endpgfextra}
	}
}

\usetikzlibrary{shadows}
\usetikzlibrary{backgrounds}
\tikzset{
	diagonal fill/.style 2 args={fill=#2, path picture={
			\fill[#1, sharp corners] (path picture bounding box.south west) -|
			(path picture bounding box.north east) -- cycle;}},
	reversed diagonal fill/.style 2 args={fill=#2, path picture={
			\fill[#1, sharp corners] (path picture bounding box.north west) |- 
			(path picture bounding box.south east) -- cycle;}}
}
\usetikzlibrary{arrows.meta}

\title{\textbf{Metastable Financial Markets}} 

\author{Diego Marcondes$^{1,2}$ and Adilson Simonis$^{2}$} 

\date{\footnotesize  $^{1}$ Department of Electrical and Computer Engineering, Texas A\&M University\\$^{2}$ Institute of Mathematics and Statistics, University of São Paulo} 

\begin{document}
	
	\maketitle 
	
	
	
	\begin{abstract}
		Metastability is a phenomenon observed in stochastic systems which stay in a false-equilibrium within a region of its state space until the occurrence of a sequence of rare events that leads to an abrupt transition to a different region. This paper presents financial markets as metastable systems and shows that, under this assumption, financial time series evolve as hidden Markov models. In special, we propose a theory that outlines an explicit causal relation between a financial market and the evolution of a financial time series. In the context of financial economics and causal factor investment, this theory introduces a paradigm shift, suggesting that investment performance fluctuations are primarily driven by the market state rather than direct causation by other variables. While not incompatible with traditional causal inference, our approach addresses the non-stationary evolution of time series through changes in market states, enhancing risk assessment and enabling mitigation strategies.
	\end{abstract}
	
	\hspace*{3,6mm}\textit{Keywords:} hidden Markov model, Markov processes, metastability, causal factor investing, financial economics 
	
	\vspace{30pt} 
	
	
	\section{Introduction}
	
	A scientific theory seeks to explain a phenomenon. Such an explanation should not only be consistent with empirical observations, but it should also be falsifiable. In the hypothetico-deductive method, the researcher, based on phenomenological observations, conjecture basic hypotheses to form a theory, and then deduce predictions which are consequences of it. In order for a theory to be scientific, the researcher should develop methods to test the predictions so, in case they are not substantiated, the theory can be regarded as false. A theory is scientific only if it is possible, in principle, to establish that it is false \cite{popper2005logic}.
	
	The deductive predictions are obtained via causal relations between the hypotheses and the observed phenomenon, which are the explanation provided by the scientific theory. Although the causal mechanism is more evident in some areas such as physics, this is not always the case, specially when interventional experimental studies are not feasible. One of these areas is factor investing, which is concerned with building investment strategies based on factors that alleged to explain differences in the performance of investments. The vast majority of research in this area identifies associations between factors and the performance of an investment, which are erroneously taken as causal relations, that cannot be falsified. Furthermore, the associations found cannot be replicated, and the investment strategies based on them have bad out-of-sample performance \cite{harvey2017presidential,harvey2019census,harvey2016and}.
	
	The paradigm of factor investment research is that the variation in the performance of an investment (i.e., time series) is caused by other variables (factors). Under this paradigm, researchers try to identify these factors in order to build optimal investment strategies. However, we argue that, under some basic reasonable assumptions, variations on the performance of an investment are not caused by other variables, but are actually caused by the interaction between market agents. In other words, market agents take actions and make decisions that cause the performance variation, and hence the \textit{market state} is its root cause. 
	
	In this paper, we propose a scientific theory consistent with the market state being the cause of investment performance variations. Based on the vast literature of successful applications of hidden Markov models to financial time series, we make four basic assumptions, one of which states that the root cause of variations in a financial time series is the market state, and then deduce that the financial time series evolves as a hidden Markov model. Specific models for the interaction between market agents can be proposed and then possibly falsified via inference methods for hidden Markov models \cite{Capp2010InferenceIH}, so the theory is scientific.
	
	More specifically, we show that, under the assumption of the existence of a market formed by agents whose states evolve on time as a metastable interacting Markov system, a financial time series evolves as a hidden Markov model. Such a metastable system stays in a false-equilibrium within a region of its state space until the occurrence of a sequence of rare events that leads to an abrupt transition to a different region. These metastable regions can be mapped into market macrostates that can be interpreted as the hidden states of a Markov chain. Under this theory, variations on the financial time series are caused by the state of the market, that is a function of the state of its agents, so the hidden Markov chain that dictates the series is actually the chain that describes the metastable behavior of the market. Hence, there is a direct stochastic causal relation between the market state and the financial time series.
	
	In Section \ref{sec_background}, we review aspects of financial economics, hidden Markov models and metastable Markov processes which are necessary to this paper, and in Section \ref{sec_main_ideas} we present the main ideas of the proposed theory for modeling financial markets. Then, in Section \ref{sec_theoreticalModel} we present the theoretical model for financial markets, in Section \ref{sec_metastability} we review the resolvent approach to metastability, and in Section \ref{sec_hidden_meta} we formally define a mathematical model compatible with that of Section \ref{sec_theoreticalModel}, linking hidden Markov models with metastability. The contents of Sections \ref{sec_main_ideas} through \ref{sec_hidden_meta} are increasing on mathematical formalism in an attempt to make this paper more accessible. In Section \ref{sec_example}, we present an example and a simulation to illustrate the mains ideas, and in Section \ref{sec_discuss} we discuss the perspectives of the proposed theory.
	
	\section{Financial Economics, Hidden Markov Models and Metastability}
	\label{sec_background}
	
	\subsection{Causality in financial economics}
	
	Financial economics is a subfield of economics concerned with trades in which money appears on both sides, specially trades of money now for money in the future, what characterizes an investment. Since the amount of a future money exchange is uncertain, these trades involve risks that should be taken into consideration in the decision to invest. A tool for reducing such risks is information, and financial economics studies the impact of information on trades involving money or, in other words, how information can aid in the decision to invest.
	
	An important approach for investing within financial economics is factor investing, in which investment decisions are made by identifying factors that explain differences in the performance of an investment, so one can make an investment properly exposed to these factors. A special kind of factors are those in which a causal relation between the factor and the return of an investment can be established. These causal factors are usually quantitative measures (variables) associated with the evolution of an asset price (time series) via econometric models. Such an approach, and a scientific culture of incentive to produce ``significant'' results, has presented many spurious ``factors'' and causal effects that do not hold in the future (are not replicable) \cite{harvey2019census,harvey2016and} and has failed to provide consistent scientific methods to investing \cite{de2023causal,harvey2017presidential}.
	
	Another possible approach for modeling causality in investing is to depart from the basic assumption that variations in, for example, an asset price are caused foremost by decisions and actions taken by market agents. These actions and decisions may be reflected on quantitative factors associated with the price variation, but, in a strict sense, what causes the price variation are the market agents and the relation between the asset price evolution and quantitative factors is actually associational and not causal.
	
\begin{figure}[ht]
		\centering
		\includegraphics[width=\textwidth]{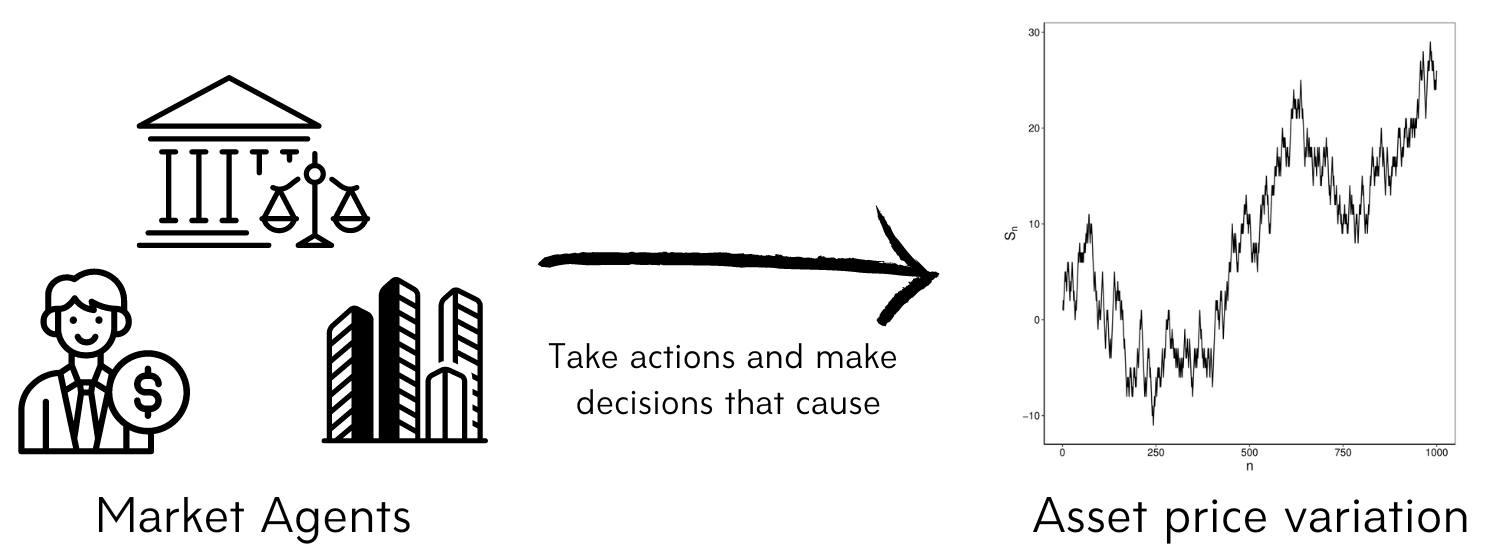}
		\caption{\footnotesize Asset price variation is caused by actions and decisions of market agents such as investors, governments, and corporations.} \label{fig_causal}
	\end{figure}

	This causal relation is illustrated in Figure \ref{fig_causal} and an approach for modeling causality in this context would consist in (a) identifying the states of market agents and (b) establishing a causal relation between these states and the asset price variation over time. This would be an approach to establishing a direct causality with the asset price variation, instead of a possible causal relation between quantitative factors and the asset price: after all, price variations are a consequence of the decisions of market agents, even if they cannot be observed and the exact causality relation may be complex.
	
	\subsection{Hidden Markov models for financial time series}
	
	Let $\{Y_{n}\}$ and $\{X_{n}\}$ be discrete-time stochastic processes with state spaces $S_{Y} = \{y_{1},\dots,y_{l}\}$ and $S_{X} = \{x_{1},\dots,x_{k}\}$, respectively. We say that the pair $(Y_{n},X_{n})$ is a hidden Markov model (HMM) if:
	\vspace{0.25cm}
	\begin{enumerate}
		\item $Y_{n}$ is a homogeneous Markov chain with transition probabilities
		\begin{linenomath}
			\begin{equation*}
				p_{i,j} \coloneqq \mathbb{P}\left(Y_{n} = y_{j} | Y_{n-1} = y_{i} \right)
			\end{equation*}
		\end{linenomath}
		for $i,j \in \{1,\dots,l\}$ and $n \geq 1$.
		\item The transition probabilities of $X_{n}$ given $Y_{n},\dots,Y_{0}$ are time-homogeneous and depend only on $Y_{n}$, that is,
		\begin{linenomath}
			\begin{equation*}
				q_{i,j} \coloneqq \mathbb{P}\left(X_{n} = x_{j} | Y_{n} = y_{i}, Y_{n-1} = y_{i_{n-1}},\dots,Y_{0} = y_{i_{0}}\right) = \mathbb{P}\left(X_{n} = x_{j} | Y_{n} = y_{i}\right),
			\end{equation*}
		\end{linenomath}
		for $j \in \{1,\dots,k\}, i,i_{n-1},\dots,i_{0} \in \{1,\dots,l\}$ and $n \geq 1$.
	\end{enumerate}
	\vspace{0.25cm} The chain $\{Y_{n}\}$ is called the hidden states of $\{X_{n}\}$ and dictates its evolution. We could also define a continuous-time HMM and a HMM with countable or continuous state space.
	
	HMM were first studied in the 1960s \cite{baum1966statistical,baum1970maximization} and have been applied in the literature to a wide range of problems in areas such as speech recognition \cite{gales2008application,jelinek1975design,juang1991hidden}, biology \cite{eddy1996hidden,krogh1994hidden} and time series \cite{macdonald1997hidden,zucchini2017hidden}, specially financial time series \cite{gupta2012stock,oelschlager2023detecting,park2009forecasting,weigend2000,zhang2019high}. 
	
	The application of HMM to financial time series is justified by the alleged existence of underlying states which determine their behavior, but are not observable. These hidden states are assumed to be related to agents of the financial market such as companies, governments and investors, whose joint states determine the market (hidden) state that evolves due to the interaction between agents. The market hidden state is assumed to influence the observable series via its transition probabilities. For instance, when a transition occurs from hidden state $y_{i}$ to $y_{i^\prime}$, the transition probabilities of $X_{n}$ change from $\{q_{i,j}: j =1,\dots,k\}$ to $\{q_{i^\prime,j}: j =1,\dots,k\}$ and hence the evolution regime of $X_{n}$ changes.
	
\begin{figure}[ht]
		\centering
		\begin{tikzpicture}
			\tikzstyle{hs} = [circle,draw=black, rounded corners,minimum width=3em, node distance=2.5cm, line width=1pt]
			\tikzstyle{hs2} = [circle,draw=black,dashed, rounded corners,minimum width=3em, node distance=2.5cm, line width=1pt]
			
			\node[hs] (ym1) at (-3,0) {-1};
			\node[hs] (y0) at (0,-2) {0};
			\node[hs] (y1) at (3,0) {1};
			
			\node[hs2] (M1xm1) at (-4,3) {-1};
			\node[hs2] (M1x1) at (-2,3) {1};
			\node[hs2] (1xm1) at (2,3) {-1};
			\node[hs2] (1x1) at (4,3) {1};
			\node[hs2] (0xm1) at (-1,-4) {-1};
			\node[hs2] (0x1) at (1,-4) {1};
			
			\node[align=center] (t) at (-6.5,-0.5) {Hidden Markov\\ chain $Y_{n}$};
			\node[align=center] (t) at (-3,4.1) {Observable states $X_{n}$};

			\begin{scope}[line width=1pt]
				\draw [decorate,decoration = {calligraphic brace}] (-5,-2) to (-5,1);
				\draw [decorate,decoration = {calligraphic brace}] (-4.5,3.65) to (-1.5,3.65);

				\draw[->] (ym1) to node[sloped,above] {0.1} (y0);
				\draw[->] (y0) to[bend left]  node[sloped,below] {0.1} (ym1);
				
				\draw[->] (ym1) to[bend left] node[sloped,above] {0.1} (y1);
				\draw[->] (y1) to node[sloped,above] {0.1} (ym1);
				
				\draw[->] (y1) to node[sloped,above] {0.1} (y0);
				\draw[->] (y0) to[bend right] node[sloped,below] {0.1} (y1);
				
				\draw[->] (y0) to[loop above]  node[sloped,above] {0.8} (y0);
				\draw[->] (y1) to[loop right]  node[right] {0.8} (y1);
				\draw[->] (ym1) to[loop left]  node[left] {0.8} (ym1);
				
				\draw[->] (y0) to node[sloped,above] {0.5} (0xm1)[dashed];
				\draw[->] (y0) to node[sloped,above] {0.5} (0x1)[dashed];
				
				\draw[->] (y1) to node[sloped,below] {0.45} (1xm1)[dashed];
				\draw[->] (y1) to node[sloped,below] {0.55} (1x1)[dashed];
				
				\draw[->] (ym1) to node[sloped,below] {0.55} (M1xm1)[dashed];
				\draw[->] (ym1) to node[sloped,below] {0.45} (M1x1)[dashed];
			\end{scope}
		\end{tikzpicture}
		\caption{\footnotesize Example of a hidden Markov model with $S_{Y} = \{-1,0,1\}$ and $S_{X} = \{-1,1\}$.} \label{dyn_hmm_example}
	\end{figure}
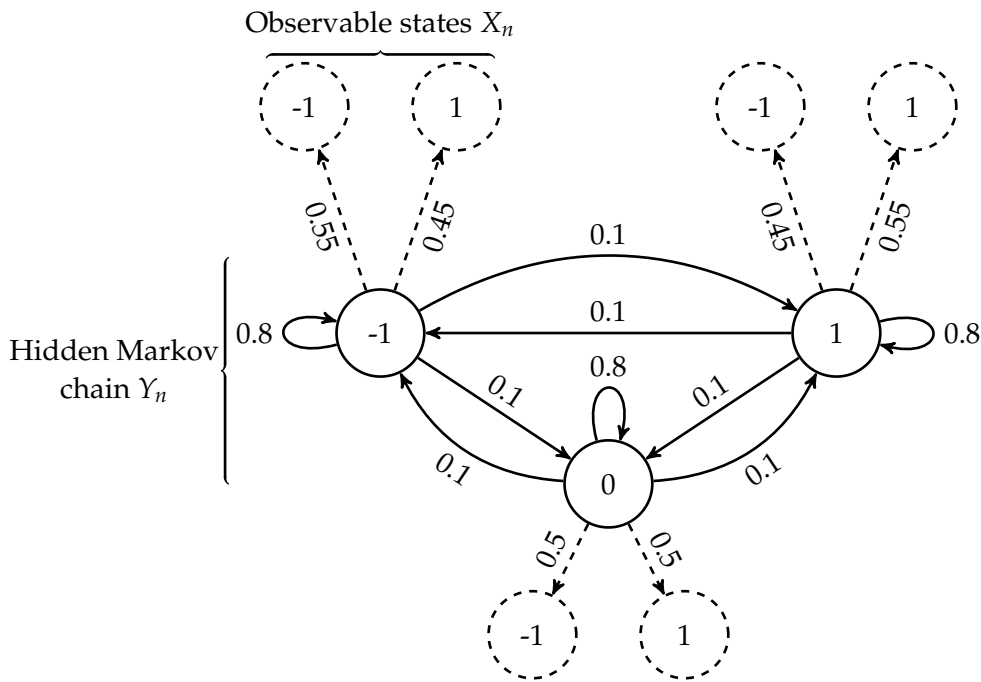

	To fix ideas, assume that $S_{Y} = \{-1,0,1\}$, $S_{X} = \{-1,1\}$ and the dynamics are those described in Figure \ref{dyn_hmm_example}. In this example, the observable chain $X_{n}$ is the variation (increase or decrease of one unit) of an asset price and the underlying states represent market conditions $1,-1$ and $0$ in which, respectively, an increase is more likely ($q_{1,1} = 0.55$), a decrease is more likely $(q_{-1,-1} = 0.55$), and both an increase and a decrease are equiprobable ($q_{0,j} = 0.5$). At each time, the hidden Markov chain $Y_{n}$ has a probability $0.8$ of staying in the same state, and a probability $0.1$ of jumping to each of the other two states. 
	
\begin{figure}[ht]
		\centering
		\includegraphics[width=\linewidth]{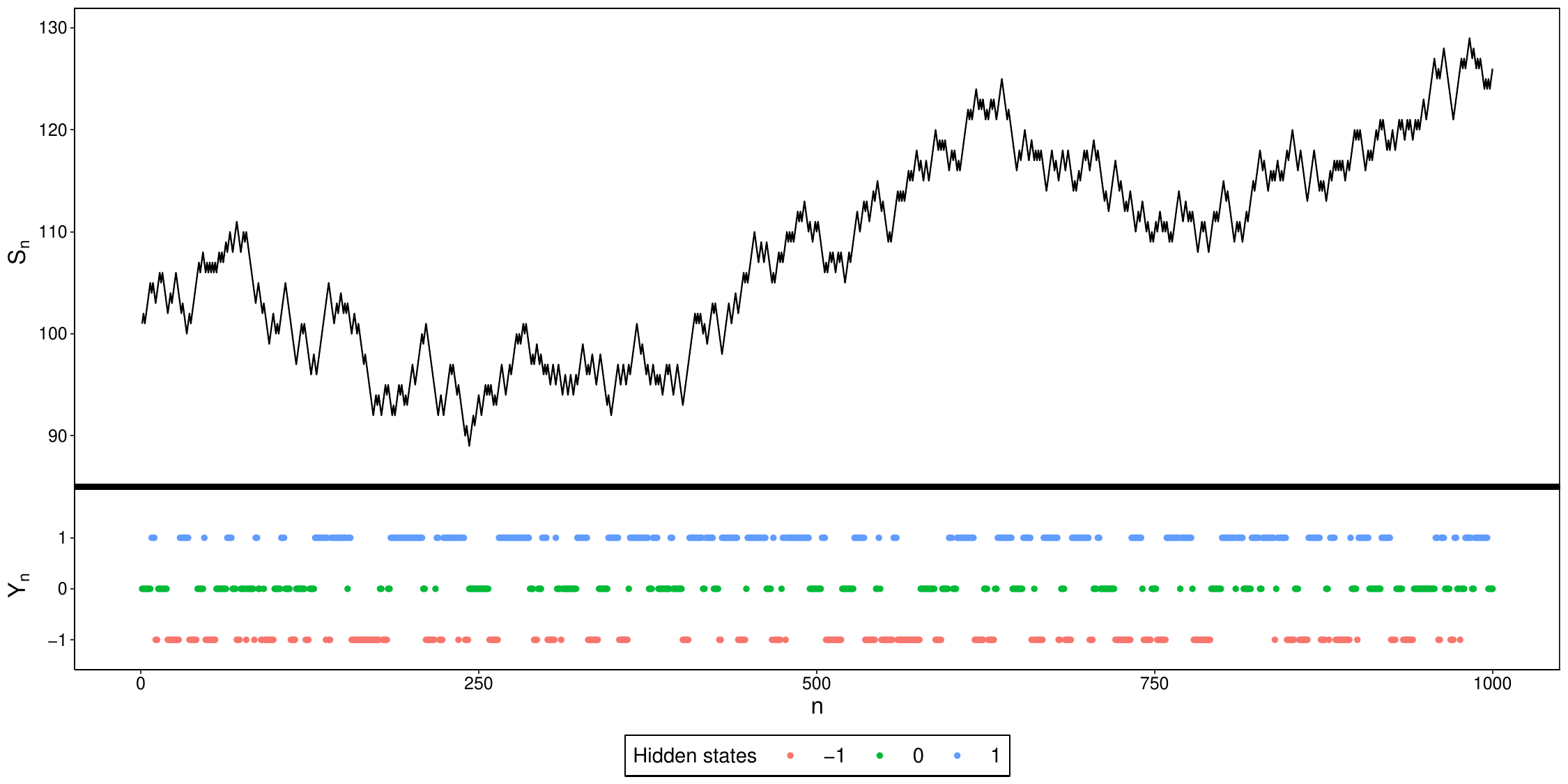}
		\caption{\footnotesize Accumulated variation $S_{n}$ (cf. \eqref{Sn}) and hidden state $Y_{n}$ over time of a simulated hidden Markov model with the dynamic described in Figure \ref{dyn_hmm_example}.} \label{fig_sim_Sn}
	\end{figure}

	In this context, consider the process $\{S_{n}\}$ given by
	\begin{linenomath}
		\begin{equation}
			\label{Sn}
			S_{n} = S_{0} + \sum_{k = 1}^{n} X_{k},
		\end{equation}
	\end{linenomath}
	that represents the price of an asset at the time $n$ when its price at the time $n = 0$ is $S_{0}$. Figure \ref{fig_sim_Sn} presents a trajectory of $S_{n}$ obtained via a simulation of the HMM with dynamic described in Figure \ref{dyn_hmm_example}. This trajectory has features of financial time series such as high volatility, periods of successive gains/losses and a trend which changes between upwards, downwards, and stable from time to time. Observe that these features are related to the hidden states since volatile periods are due to a rapid change between hidden states, successive gains/losses are due to the persistence of a hidden state and the trend changes are due to a change in the time spent on each hidden state in the recent past.
	
	This simple example illustrates that features present in financial time series could in principle be represented by a HMM and, since the existence of hidden states determining the series dynamic is a reasonable practical assumption in this instance, the modeling of financial time series by HMM is in principle justifiable. We refer to \cite{gupta2012stock,macdonald1997hidden,oelschlager2023detecting,park2009forecasting,weigend2000,zhang2019high,zucchini2017hidden} and the references therein for more sophisticated models and more details on the statistical inference of HMM.
	
	\subsection{Metastability of Markov processes}
	
	The study of the metastability of Markov processes has been motivated by a homonym physical phenomenon. The first rigorous method for deducing the metastable behavior of Markov processes, called \textit{pathwise approach to metastability}, was proposed in the 1980s by \cite{cassandro1984metastable}, and was successfully applied in the following decades to many models in statistical mechanics \cite{bianchi2020soft,durrett1988contact,fernandez2015asymptotically,galves1987metastability,mathieu1998metastability,mountford1993metastable,olivieri2005large,peixoto1995metastable,schonmann1991approach,simonis1996metastability}. 
	
	This approach considers that metastability is a phenomenon occurring when a process ``has a unique stationary probability measure, but if the initial conditions are suitably chosen then the time to get to the asymptotic state becomes very large, and during this time the system behaves as if it were described by another stationary measure; finally, and abruptly, it goes to the true equilibrium'' \cite{cassandro1984metastable}. In special, the pathwise approach considers single typical trajectories and studies their averages to detect if the time averages over almost all trajectories are practically stationary for a very long period, until some large fluctuation leads the system to another, entirely different situation.
	
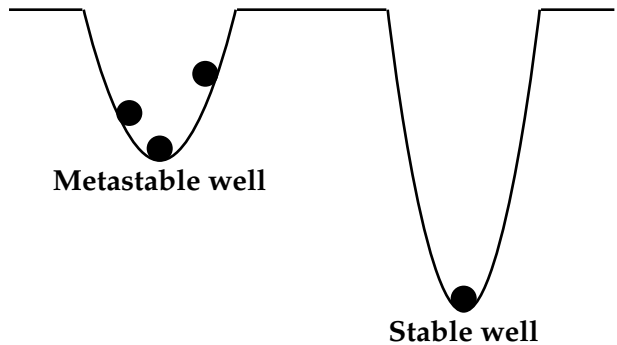
\begin{figure}[ht]
		\centering
		\begin{tikzpicture}[every node/.style={inner sep=0,outer sep=0}]
			\tikzstyle{hs} = [circle,fill,draw=black, rounded corners,minimum width=0.8em, node distance=2.5cm, line width=1pt]
			
			\node (p1) at (-3,0) {};
			\node (p2) at (-2,0) {};
			\node (p3) at (-1,-2) {};
			\node[hs] (p3Ball) at (-1,-1.84) {};
			\node[hs] (p35Ball) at (-0.4,-0.85) {};
			\node[hs] (p25Ball) at (-1.4,-1.37) {};
			\node (p3Text) at (-1,-2.25) {\textbf{Metastable well}};
			\node (p4) at (0,0) {};
			\node (p5) at (2,0) {};
			\node (p6) at (3,-4) {};
			\node (p6Text) at (3,-4.25) {\textbf{Stable well}};
			\node[hs] (p6Ball) at (3,-3.83) {};
			\node (p7) at (4,0) {};
			\node (p8) at (5,0) {};
			
			\draw [line width=1pt] plot [smooth, tension=1] coordinates { (-2,0) (-1,-2) (0,0)};
			\draw [line width=1pt] plot [smooth, tension=1] coordinates {(2,0) (3,-4) (4,0)};
			
			\begin{scope}[line width=1pt]
				\draw[-] (p1) -- (p2);
				\draw[-] (p4) -- (p5);
				\draw[-] (p7) -- (p8);
			\end{scope}
			
		\end{tikzpicture}
		\caption{\footnotesize Illustration of the metastability phenomenon studied by the pathwise approach to metastability. In this characterization of metastability, the process spends an exponential time in a metastable well, as if it had attained stability, until a sequence of rare events causes the process to abruptly leave this well and attain its real stability in another deeper well. The depth of the well is associated with the timescale in which the process attains the equilibrium inside the well.} \label{fig_pathwise}
	\end{figure}

	Figure \ref{fig_pathwise} illustrates the metastable behavior described by the pathwise approach, in which the process, when starting from a state in a so-called metastable well, remains an exponential time inside the well, until a sequence of rare events takes the process outside this well, and then abruptly into the effective stable well, where the process will remain. The pathwise approach studies the average of typical trajectories to determine the order of this exponential time and the likely paths the process takes from, for example, the bottom of the metastable well to the stable well. If the process is accelerated by this time order, then one will observe the metastable behavior; if the process is accelerated by a time of lesser order, then it may not escape the metastable well and other specific behaviors may be observed (for example the nucleation phase of the process \cite{beltran2017martingale}); and if the process is accelerated by a time of greater order, then it will instantaneously attain the stable well and no metastable behavior is observed.
	
	In the 2000s, a new approach for metastability based on potential theory was proposed by \cite{bovier2002metastability,bovier2004metastability,gayrard2005metastability}. In the \textit{potential theoretic approach to metastability}, large deviations tools are replaced with potential theory to derive sharp estimates for the transition times between wells. It differs from the pathwise approach since, instead of identifying the most likely paths and estimating their probabilities, in the potential theoretic approach the metastability phenomenon is interpreted as a sequence of visits to different metastable wells and the approach focuses on the precise analysis of the respective hitting probabilities and hitting times of these wells with the help of potential theory, e.g, by solving suitable Dirichlet problems. A detailed account of this approach to metastability can be found in \cite{bovier2016metastability}.
	
	More recently, a third approach for metastability was proposed through the characterization of Markov processes as unique solutions of martingale problems \cite{beltran2015martingale}. In the \textit{martingale approach to metastability}, the metastability phenomenon is interpreted similarly as in the potential theoretic approach, but besides the hitting times of metastable wells, metastability is also described by a simple Markov chain that dictates the time evolution of the wells visits. In this model, there are $\kappa \geq 2$ metastable wells with the same depth and, starting from a metastable well, the process spends an exponential time in it before jumping to another metastable well, spending a negligible amount of time outside the metastable wells. 
	
	Formally, metastability is derived by studying the limit of the order process, that is the process with state space $S = \{1,\dots,\kappa\}$ which represents the metastable well where the trace process is at each time. The trace process is obtained from the original process by turning off the clock when the process is not in a metastable well. The existence of such Markov process limit of the order process has been known in the literature as \textit{tunneling}, while the word metastability used to be employed exclusively to characterize instances in which there are also stable wells that are eventually reached. However, more recently, the word metastability has been widely employed to characterize tunneling as well.
	
\begin{figure}[ht]
		\centering
		\begin{tikzpicture}[every node/.style={inner sep=0,outer sep=0}]
			\tikzstyle{hs} = [circle,fill,draw=black, rounded corners,minimum width=0.8em, node distance=2.5cm, line width=1pt]
			
			\node (p1) at (-3,0) {};
			\node (p2) at (-2,0) {};
			\node (p3) at (-1,-2) {};
			\node (p3Text) at (-1,-2.25) {$\boldsymbol{\mathscr{E}_{N}^{1}}$};
			\node (p3Y) at (-1,0.25) {$Y_{N}(t) = 1$};
			\node (p4) at (0,0) {};
			\node (pDelta) at (1,0.25) {$\boldsymbol{\Delta_{N}}$};
			\node (p5) at (2,0) {};
			\node (p6) at (3,-2) {};
			\node (p6Text) at (3,-2.25) {$\boldsymbol{\mathscr{E}_{N}^{2}}$};
			\node (p6Y) at (3,0.25) {$Y_{N}(t) = 2$};
			\node (p7) at (4,0) {};
			\node (p71) at (5,0) {};
			\node (p8) at (6,-2) {};
			\node (p8Text) at (6,-2.25) {$\boldsymbol{\mathscr{E}_{N}^{3}}$};
			\node (p8Y) at (6,0.25) {$Y_{N}(t) = 3$};
			\node (p9) at (7,0) {};
			\node (p10) at (8,0) {};
			
			\draw [line width=1pt,blue] plot [smooth, tension=1] coordinates { (-2,0) (-1,-2) (0,0)};
			\draw [line width=1pt,blue] plot [smooth, tension=1] coordinates {(2,0) (3,-2) (4,0)};
			\draw [line width=1pt,blue] plot [smooth, tension=1] coordinates {(5,0) (6,-2) (7,0)};
			
			\begin{scope}[line width=1pt]
				\draw[-] (p1) -- (p2);
				\draw[-] (p4) -- (p5);
				\draw[-] (p7) -- (p71);
				\draw[-] (p9) -- (p10);
			\end{scope}
			
		\end{tikzpicture}
		\caption{\footnotesize Illustration of metastability in the sense of tunneling. In this instance, there are three wells with the same depth and the order process $Y_{N}(t)$ represents the well the trace process is at time $t$. Metastability occurs when the process $Y_{N}(t)$ converges to a Markov process, and the original process spends a negligible amount of time outside the metastable wells, that is, in $\Delta_{N}$, when $N$ tends to infinite.} \label{fig_tunneling}
	\end{figure}
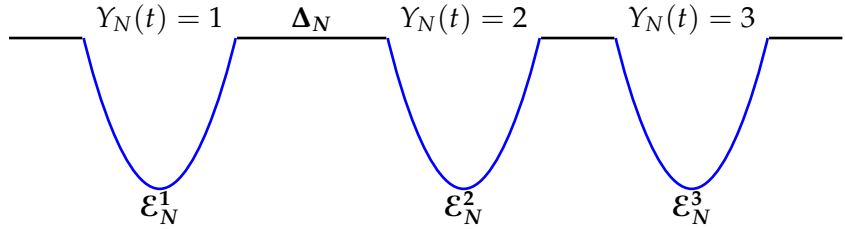

	Metastability in the sense of tunneling is illustrated in Figure \ref{fig_tunneling}. In this case, we assume there is a sequence of Markov processes $\{\{\xi_{N}(t)\}: N \geq 1\}$ and that, for each $N$, there are three wells with the same depth. The order process $Y_{N}(t)$ represents the well the trace process of $\xi_{N}(t)$ is at time $t$. Metastability occurs when the order process $Y_{N}(\cdot)$ converges to a Markov process, and the time $\xi_{N}(\cdot)$ spends outside the metastable wells, that is, in $\Delta_{N}$, converges to zero, when $N$ tends to infinity.
	
	Through the martingale characterization of Markov chains, sufficient conditions for the convergence of the order process to a Markov process were derived based on potential theory by assuming the existence of an attractor inside each well \cite{beltran2010tunneling,beltran2012tunneling}, i.e., a state that is visited with probability tending to one before the process leave the respective well. Sufficient conditions based on the mixing time and on the relaxation time of the process reflected at the boundary of the metastable wells were also derived \cite{beltran2015martingale}. The martingale approach has been successfully applied, and we refer to \cite{landim2019metastable} for a review.
	
	Since the sufficient conditions for metastability derived from the martingale approach usually demand the existence of an attractor, sharp estimates of the mixing time of the reflected process and explicit knowledge of the stationary distribution of the process, they may not be proved to some processes, including weakly mixing Markov processes. Motivated by the metastability of the critical zero-range process, on which occur a phase transition on the metastable behavior established for the super-critical case in \cite{beltran2012metastability} and for which the sufficient conditions of \cite{beltran2012tunneling,beltran2010tunneling,beltran2015martingale} are not satisfied, a new approach for metastability based on the solution of resolvent equations was proposed by \cite{landim2023metastable}. The \textit{resolvent approach} to metastability, as showed by \cite{landim2021resolvent}, provides not only sufficient, but also necessary conditions for metastability based on properties of the solution of resolvent equations, which are implied by the conditions in \cite{beltran2012tunneling,beltran2010tunneling,beltran2015martingale}, but that may in principle also be showed for processes that do not satisfy them. Applications of the resolvent approach for metastability are ongoing, but it has already been successfully applied \cite{kim2023hierarchical,landim2023metastability,lee2022non}.
	
	Metastability, in the sense of tunneling\footnote{From now on, when we say metastability we mean in the sense of tunneling and not as in the pathwise approach.}, occurs when a macro behavior of the process can be described in the limit by a Markov chain. Although the state space of the process may contain states with specific characteristics, there are some clusters of states with common features, and such that the process spends a negligible time outside them. In this instance, the macroevolution of the process can be described by the dynamic of its visits to these clusters by a Markov chain.
	
	\section{Main ideas of a theory for modeling financial markets}
	\label{sec_main_ideas}
	
\begin{figure}[ht]
		\centering
		\includegraphics[width=\textwidth]{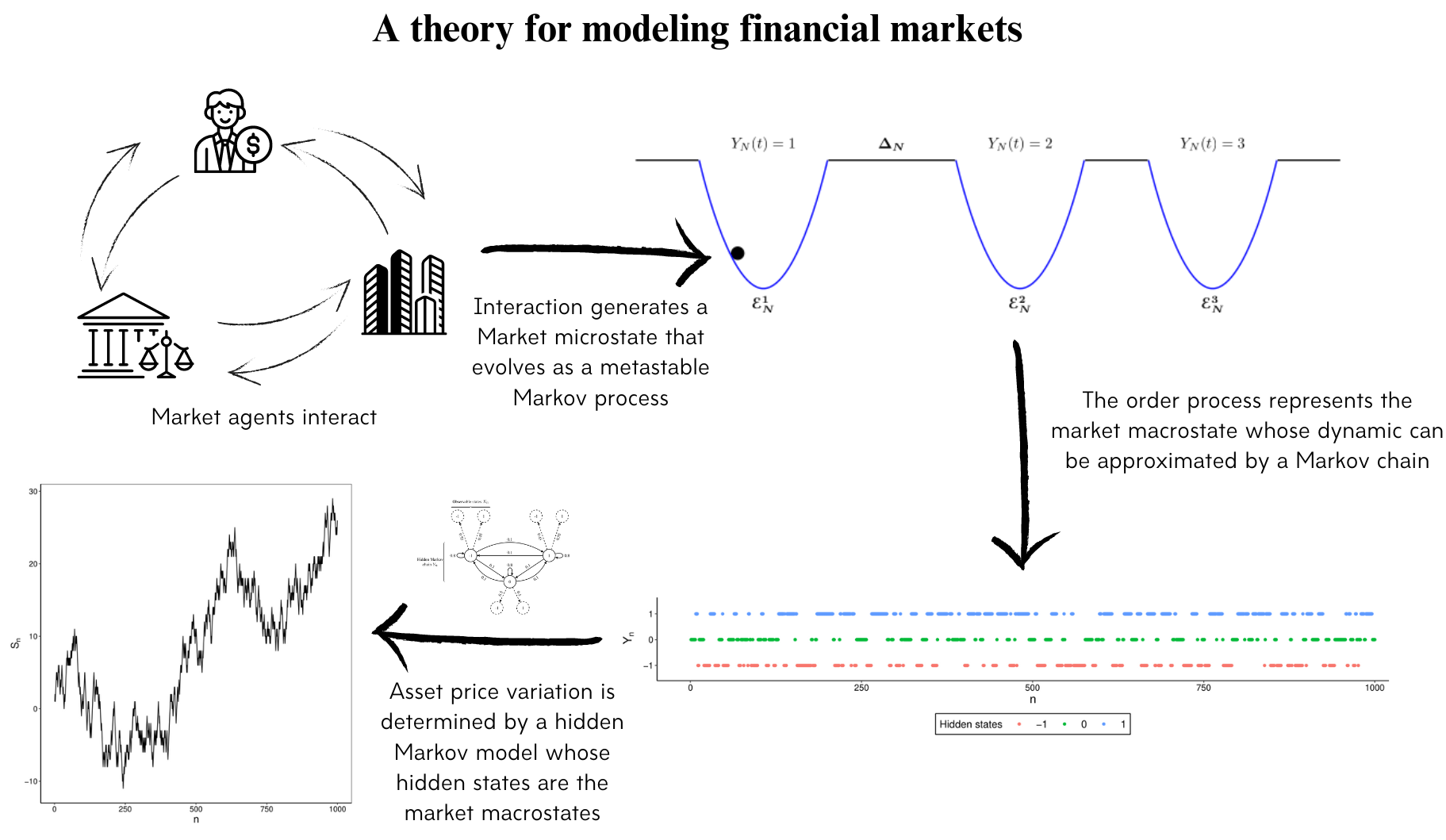}
		\caption{\footnotesize Summary of a theory for modeling financial markets.} \label{fig_main_diagram}
	\end{figure}

	The theory proposed in this paper combines the ideas presented in Section \ref{sec_background} to deduce that hidden Markov models in finance are implied by the existence of a market formed by agents whose states evolve on time as a metastable interacting Markov system. This system is akin to interacting particle systems \cite{liggett1985interacting}, but the ``particles'' are replaced by market agents which interact over time, generating a market microstate that is a function of its agents states. As a consequence of metastability, the market can be described by macrostates and the dynamic of the visits to them can be approximated by a Markov chain. In turn, these macrostates are the hidden states of a HMM, which determine the variation of an observable financial time series over time. The HMM represents a direct causal relation between the market state and the time series. This framework is summarized in Figure \ref{fig_main_diagram}.
	
	The theory is deduced from four assumptions that, informally, state:
	\begin{enumerate}
		\item[\textbf{A1}] A market is formed by interacting agents, and the state of the market is a function of the agents states and evolves as a Markov process;
		\item[\textbf{A2}] The timescale on which the agents interact is lesser than the timescale on which financial time series are observed;
		\item[\textbf{A3}] The market Markov process speeded-up to the time series timescale is metastable;
		\item[\textbf{A4}] The transition probabilities of the observable time series depend on the market state mainly through its macrostate. In other words, the transition probabilities are approximately equal for all microstates within a same metastable well (macrostate).
	\end{enumerate}
	
	The actions and decisions of an agent lead to its \textit{state}. For example, if these actions and decisions are consistent with the expectation that an asset price will increase, its state is $1$; if they are consistent with a future decrease in the price, its state is $-1$; and if they are consistent with a stabilization of the price, its state is $0$. A change of state is caused by an interaction with other market agents and their states, which may cause the agent to change its expectations about future prices. The market state is a function of its agents states and the evolution of the market depends on the past only through the current state, so it is a Markov process. This is assumption \textbf{A1}.
	
	Since the market agents interact in \textit{real-time}, and financial time series are observed in a specific timescale (minutes, hours, days, etc...), the timescale in which agents interact is lesser than that of observable time series. The fact that market agents are making decisions and taking actions \textit{at all times}, so the market state is constantly changing, can be better understood by noting that there are a huge number of market agents and, at any given time, some of them are changing their state, and also by noting that many market agents are actually represented by algorithms which take actions in the timescale of milliseconds or less, in what is known as high-frequency trading \cite{biais2011high,jones2013we}. These facts are assumption \textbf{A2}.
	
	The metastable behavior of financial markets assumed by \textbf{A3} is supported by empirical evidence and historical experiences. As an illustration, let's analyze how some transitions have occurred in the past between a \textit{bull} and a \textit{bear} market, characterized, respectively, by a general increase and decrease in the market value of assets. We discuss two recent historical events:
	\begin{itemize}
		\item \textbf{2007-2008 financial crisis}: The financial crisis started in the United States with the bursting of a housing bubble and the growth of mortgage defaults. In summary, it was a consequence of actions and decisions that market agents took in the preceding years concerning subprime mortgages products that had been extended in growing numbers at the height of the bubble to less creditworthy borrowers, although many other factors contributed to the crisis \cite{helleiner2011understanding}. Before the crisis, a stabilized bull market had endured since the recover from the \textit{2000-2002 dotcom crash} \cite{lieberman2005did}, until a sequence of events, in the US and then worldwide, led to an abrupt (matter of weeks) end of this bull market and, after government bail-outs, the stabilization within a new market macrostate \cite{wilson2012fiscal}. In this example, the financial crisis represents the abrupt transition between two metastable wells over which the macrostate of the market is different. Observe that this abrupt change was caused by actions taken by the market agents.
		
		\item \textbf{COVID-19 pandemic}: The stable bull market attained after the recuperation from the 2007-2008 financial crisis ended abruptly in March 2020 due to the COVID-19 pandemic \cite{ceylan2020historical,jackson2021global}. The economic effects of containment measures \cite{deb2022economic} and the spontaneous change of behavior of people due to the pandemic have entirely changed the global market microstate and a different stabilized market, the \textit{new normal} \cite{buheji2020planning}, emerged. The pandemic may be seen as an abrupt change between metastable wells caused by a factor exogenous to the market, but that influences it.
	\end{itemize}
	
	Although both examples above are related to the global market and how it affects economic indicators as a whole, the same underlying ideas ought to be applied to more specific markets. Significant changes in the market state do not occur slowly and gradually since, otherwise, market agents that benefit from the current state of affairs would identify this change and act to avoid it, while others would act to profit from it, and these actions would cause such a change to probably not happen at all. Rather, these changes are abrupt and mainly unexpected, ending an apparent stability, i.e., metastability, and starting a new one. The examples also outline that the market microstate transition rates may not depend deterministically on the current microstate, but may also comprise a term that accounts for random transitions due to exogenous uncontrolled factors such as pandemics and geopolitical conflicts.
	
	Finally, assumption \textbf{A4} is supported by the fact that, in general, a market agent cannot by itself cause a variation on a financial time series. These variations are caused by a feature of the joint state of the agents, i.e., a macrostate. The case in which there are powerful agents, such as governments, which can take direct (with a stroke of a pen) or indirect actions to cause a variation in a time series, may also be framed within this assumption, since not all agents may have the same weight on the market state. In summary, the causality relation between the market state and a financial series is through general and not specific features of the market.
	
	Denote by $X(t)$ the continuous-time observed time series and by $Y(t)$ the Markov process that approximates the order process that describes the macrostate of the market when its Markov process is speeded-up to be in the same timescale as the financial series. 
	
	The main result of this paper states that, under assumptions \textbf{A1}-\textbf{A4}, $(X(t),Y(t))$ is a hidden Markov model. In other words, we show that the modeling of financial time series by HMM is theoretically justified under the assumption that the market evolves as a metastable interacting Markov system and determines the transition rates of the financial time series through its macrostate. In the next section, we describe in more details this proposed theoretical model for financial markets.
	
	\section{A theoretical model for metastable financial markets}
	\label{sec_theoreticalModel}
	
	A financial market $\mathcal{M}_{N} = \{a_{1},\dots,a_{N}\}$ is a set of $N > 1$ \textit{market agents}. At any given moment, the state of an agent is one of the states in a finite set $S$ and the market state is a function of the state of its agents. We denote the collection of possible market states by $\mathcal{H}_{N}$. For example, we could have $S = \{-1,1\}$ and $\mathcal{H}_{N} = \{-1,1\}^{\mathcal{M}_{N}}$, so each agent is in one of two states and the market state is the joint states of its agents. Another example would be $S = \{1,\dots,\kappa\}$ and
	\begin{linenomath}
		\begin{equation*}
			\mathcal{H}_{N} = \left\{\eta = (\eta_{1},\dots,\eta_{\kappa}) \in \mathbb{\mathbb{Z}}_{+}^{\kappa}: \sum_{i=1}^{\kappa} \eta_{i} = N\right\},		
		\end{equation*}
	\end{linenomath}
	in which each agent can be in one of $\kappa$ states and the market state is the number $\eta_{i}$ of agents in each of the $\kappa$ states. Observe that in the first example the state of each agent is known from the market state, while in the second example it is known from the market state how many agents are in each state, but not which agents are in each state.
	
	The time evolution of the market state is described by an interacting Markov system, in which each agent changes its state at a rate depending on the state of other agents. We define $\eta_{N}: \mathbb{R}_{+} \to \mathcal{H}_{N}$ as the time evolution of the market, in which $\eta_{N}(t)$ is the state of the market at a time $t \geq 0$. We assume that the state of the market $\eta_{N}(\cdot)$ evolves as an irreducible, continuous-time Markov process, as summarized in the first model assumption.
	
	\vspace{0.25cm}
	\begin{enumerate}
		\item[] \textbf{Assumption A1}: A market is formed by interacting agents, and the state of the market $\eta(\cdot)$ is a function of the agents states and evolves as a $\mathcal{H}_{N}$-valued, irreducible, continuous-time Markov process.
	\end{enumerate}
	\vspace{0.25cm}
	
	Within a market, there are many quantities of interest that are observed over time which, although depend on the market state, are not a deterministic function of it, since there may be exogenous random factors that also influence them. An example of such quantities is the variation of asset prices, currencies rates and price indexes. We denote the time evolution of such a quantity by $X_{N}: \mathbb{R}_{+} \to \mathcal{X}$, so $X_{N}(t)$ represents the quantity, that take values in a set $\mathcal{X}$, at a time $t \geq 0$. In order to ease the presentation of the main ideas of this paper, we assume that $\mathcal{X}$ is a countable set. We call $X_{N}(\cdot)$ a financial time series.
	
	While the market agents interact in \textit{real-time}, these quantities of interest are observed on another timescale, for example, every minute, every hour or every day. We denote the ratio between the time-scales of $X_{N}(\cdot)$ and $\eta_{N}(\cdot)$ by $\theta_{N}$. We assume that $\theta_{N}$ depends on the number $N$ of market agents and that it diverges when $N$ tends to infinity. In order for the process $\eta_{N}(\cdot)$ to be in the same timescale as $X_{N}(\cdot)$ it should be speeded-up by $\theta_{N}$. Defining the speeded-up process $\xi_{N}(t) \coloneqq \eta(\theta_{N} t)$, the processes $\xi_{N}(\cdot)$ and $X_{N}(\cdot)$ are on the same timescale. This is the second model assumption.
	
	\vspace{0.25cm}
	\begin{enumerate}
		\item[] \textbf{Assumption A2}: The timescale on which the agents interact is lesser than the timescale on which financial time series are observed. The ratio $\theta_{N}$ between the time-scales depends on $N$ and $\lim\limits_{N \to \infty} \theta_{N} = \infty$.
	\end{enumerate}
	\vspace{0.25cm}
	
	When the number $N$ of agents tends to infinity, the speeded-up process $\xi_{N}$ may be metastable, in the sense that the visits to the macrostates of the market can be approximated by a simple Markov chain. For example, for $S = \{-1,+1\}$ and $\mathcal{H}_{N} = \{-1,+1\}^{\mathcal{M}_{N}}$, this could mean that the process $\xi_{N}(\cdot)$ spends an exponential time on the subset of configurations on which all, but a finite number of agents, are in state $-1$, and then jumps to the subset of configurations on which all, but a finite number of agents, are in state $+1$ (and then vice versa). These two subsets of configurations are the metastable wells of $\xi_{N}(\cdot)$ defined as
	\begin{linenomath}
		\begin{align*}
			\mathscr{E}_{N}^{-1} &= \left\{\eta \in \mathcal{H}_{N}: \sum_{i=1}^{N} \mathds{1}\{\eta_{a_{i}} = -1\} \geq N - \ell\right\}\\
			 \mathscr{E}_{N}^{+1} &= \left\{\eta \in \mathcal{H}_{N}: \sum_{i=1}^{N} \mathds{1}\{\eta_{a_{i}} = +1\} \geq N - \ell\right\},
		\end{align*}
	\end{linenomath}
	for a $\ell \in \mathbb{Z}_{+}$ fixed. Denoting $\mathscr{E}_{N} = \mathscr{E}_{N}^{-1} \cup \mathscr{E}_{N}^{+1}$ we have that $\Delta_{N} = \mathcal{H}_{N}\setminus \mathscr{E}_{N}$ represents the configurations outside the metastable wells.
	
	The macrostate of the market is the state (metastable well) in $S$ where the great majority of the agents are at a given time and, under the assumption that the process $\xi_{N}(\cdot)$ spends a negligible amount of time in $\Delta_{N}$, a metastable behavior would mean that the transition between the macrostates $S$ occur according to a Markov process when $N$ tends to infinity. This is the third model assumption.
	
	\vspace{0.25cm}
	\begin{enumerate}
		\item[] \textbf{Assumption A3}: The speeded-up market Markov process $\xi_{N}$ is meta- stable.
	\end{enumerate}
	\vspace{0.25cm}
	
	Assumption \textbf{A3} is justifiable, since financial markets tend to have periods of relative stability which end abruptly, and then a new regime starts. In the example above, if the states $S = \{-1,+1\}$ are seen as the \textit{opinion} of each agent regarding some aspect of the market, then the metastable behavior would represent the \textit{consensus} of the market about such aspect, that may change abruptly from time to time. If this aspect is the future price of an asset, that may go up ($+1$) or down ($-1$), then a metastable market would be such that a consensus about the future variation of the asset price remains for an exponential time, until it reverts in a negligible amount of time, and a new consensus holds for another exponential time.
	
	Denote by $Z_{N}(t) = (\xi_{N}(t),X_{N}(t))$ the joint process of $\xi_{N}(\cdot)$ and $X_{N}(\cdot)$. We assume $Z_{N}(\cdot)$ is an irreducible, continuous-time Markov process and that the jumping rates of $X_{N}(\cdot)$ depend solely on $\xi_{N}(\cdot)$ and not on the current state of $X_{N}(\cdot)$. That is, the process $Z_{N}(\cdot)$ jumps from, say, $(\xi,x_{1})$ to $(\xi,x_{2})$ with a rate that depends on $\xi$ and $x_{2}$, but not on $x_{1}$. This means that, although $Z_{N}(\cdot)$ and $\xi_{N}(\cdot)$ are Markov processes, $X_{N}(\cdot)$ is not Markov, but its evolution is dictated by the hidden Markov process $\xi_{N}(\cdot)$, so $Z_{N}(\cdot)$ is actually a HMM. Furthermore, we assume that, as $N$ tends to infinity, the process $X_{N}(\cdot)$ tends to a process $X(\cdot)$ with jumping rates constant inside each metastable well of $\xi_{N}(\cdot)$. This is the last model assumption.
	
	\vspace{0.25cm}
	\begin{enumerate}
		\item[] \textbf{Assumption A4}: The process $Z_{N}(\cdot)$ is an irreducible, continuous-time Markov process with jumping rates from $(\xi,x_{1})$ to $(\xi,x_{2})$ depending on $\xi$ and $x_{2}$, but not on $x_{1}$, implying that the evolution of $X_{N}(\cdot)$ is dictated by the hidden Markov process $\xi_{N}(\cdot)$ and $Z_{N}(\cdot)$ is a HMM. Furthermore, as $N$ tends to infinity, the jumping rates from $(\xi,x_{1})$ to $(\xi,x_{2})$ are constant inside each metastable well of $\xi_{N}(\cdot)$.
	\end{enumerate}
	\vspace{0.25cm}
	
	On the one hand, this last assumption implies that the evolution of the financial time series is dictated (caused) by the state of the market, so $\xi_{N}(\cdot)$ has an effect on the financial time series $X_{N}(\cdot)$ and fluctuations on the time series are caused by the state of the market. On the other hand, the financial time series may not be a deterministic function of the market state, and hence it cannot be completely determined by the market state. This assumption is plausible, since the major cause of financial time series variation are decisions taken by market agents, i.e., the market state, but there is a random factor in the series variation that cannot be explained by the market.
	
	The main result of this paper is that, under assumptions \textbf{A1, A2, A3} and \textbf{A4}, the financial time series $X(\cdot)$, that is the series obtained from $X_{N}(\cdot)$ by making $N$ goes to infinity, which is the time series actually observed, is dictated by the hidden Markov chain $Y(\cdot)$ that describes the metastable behavior of the market. This implies that the hypothesis that a hidden Markov chain dictates the evolution of a financial time series is a consequence of a theory that describes the interaction between market agents as metastable interacting Markov systems.
	
	\begin{theorem}
		\label{theorem_intro}
		Under assumptions \textbf{A1, A2, A3} and \textbf{A4}, the evolution of the time series $X(\cdot)$ is dictated by the hidden Markov chain $Y(\cdot)$ and $(X(\cdot),Y(\cdot))$ is a hidden Markov model.
	\end{theorem}
	
	In the next section, we formally define the metastability of Markov processes, and in Section \ref{sec_hidden_meta} we propose a model compatible with assumptions \textbf{A1} to \textbf{A4}.
	
	\section{Metastability of Markov processes}
	\label{sec_metastability}
	
	The following characterization of metastability of continuous-time Markov processes was proposed by \cite{beltran2010tunneling} and is based on the martingale representation of Markov processes \cite{landim2019metastable}. We refer to \cite{landim2021resolvent} for more details.
	
	\subsection{Model definition}
	
	Let $\{\mathcal{H}_{N}: N \in \mathbb{Z}_{+}\}$ be a sequence of countable sets, and let $\{(\xi_{N}(t): t \geq 0): N \in \mathbb{Z}_{+}\}$ be a collection of $\mathcal{H}_{N}$-valued, irreducible, continuous-time Markov chains with generator $\mathcal{L}_{N}$, which acts on functions $F: \mathcal{H}_{N} \to \mathbb{R}$ as 
	\begin{linenomath}
		\begin{equation}
			\label{generatorLN}
			(\mathcal{L}_{N}F)(\eta) = \sum\limits_{\xi \in \mathcal{H}_{N}} \mathfrak{R}_{N}(\eta,\xi) \left[F(\xi) - F(\eta)\right], \eta \in \mathcal{H}_{N},
		\end{equation}
	\end{linenomath}
	where $\mathfrak{R}_{N}: \mathcal{H}_{N} \times \mathcal{H}_{N} \to \mathbb{R}_{+}$ are the jump rates. The Markov process generated by $\mathcal{L}_{N}$ is such that a jump from $\eta$ to $\xi$ occurs at a rate $\mathfrak{R}_{N}(\eta,\xi)$.
	
	Let $S$ be a finite set and $\mathscr{E}_{N}^{s}, s \in S$, be a family of disjoint subsets of $\mathcal{H}_{N}$. Denote
	\begin{linenomath}
		\begin{align*}
			\mathscr{E}_{N} = \bigcup\limits_{s \in S} \mathscr{E}_{N}^{s} & & \text{ and } & & \Delta_{N} = \mathcal{H}_{N} \setminus \mathscr{E}_{N}.
		\end{align*}
	\end{linenomath}
	The sets $\mathscr{E}_{N}^{s}$, called \textit{wells}, represent the metastable sets of the dynamics $\xi_{N}(\cdot)$ in the sense that, once the process enters one of these sets $\mathscr{E}_{N}^{s}$, it equilibrates inside it before hitting a new set $\mathscr{E}_{N}^{r}, r \neq s$. The goal of the theory is to describe the evolution between these sets.
	
	In order to accomplish this, we consider the order process, that represents in which $\mathscr{E}_{N}^{s}$ the trace process on $\mathscr{E}_{N}$ is at each time, hence represents the evolution dynamic between the sets $\mathcal{E}_{N}^{s}$.
	
	\subsection{Order process}
	
	For $\mathscr{A} \subset \mathcal{H}_{N}$, denote by $T^{\mathscr{A}}(t)$ the total time the process $\xi_{N}(\cdot)$ spends in $\mathscr{A}$ in the time-interval $[0,t]$:
	\begin{linenomath}
		\begin{equation*}
			T^{\mathscr{A}}(t) = \int_{0}^{t} \chi_{\mathscr{A}}(\xi_{N}(s)) \ ds
		\end{equation*}
	\end{linenomath}
	in which $\chi_{\mathscr{A}}$ is the characteristic function of set $\mathscr{A}$. Denote by $S^{\mathscr{A}}(t)$ the generalized inverse of $T^{\mathscr{A}}(t)$:
	\begin{linenomath}
		\begin{equation*}
			S^{\mathscr{A}}(t) = \sup\{s \geq 0: T^{\mathscr{A}}(s) \leq t\}.
		\end{equation*}
	\end{linenomath}
	
	The trace of $\xi_{N}(\cdot)$ on $\mathscr{A}$, denoted by $(\xi_{N}^{\mathscr{A}}(t): t \geq 0)$, is defined by
	\begin{linenomath}
		\begin{equation*}
			\xi_{N}^{\mathscr{A}}(t) = \xi_{N}(S^{\mathscr{A}}(t)), t \geq 0,
		\end{equation*}
	\end{linenomath}
	that is a $\mathscr{A}$-valued, continuous-time Markov chain, obtained by turning off the clock when process $\xi_{N}(\cdot)$ visits set $\mathscr{A}^{c}$.
	
	Let $\Psi_{N}: \mathscr{E}_{N} \to S$ be the projection given by
	\begin{linenomath}
		\begin{equation*}
			\Psi_{N}(\eta) = \sum_{s \in S} s \chi_{\mathscr{E}_{N}^{s}}(\eta).
		\end{equation*}
	\end{linenomath}
	The order process $(Y_{N}(t): t \geq 0)$ is defined as
	\begin{linenomath}
		\begin{equation*}
			Y_{N}(t) = \Psi_{N}(\xi_{N}^{\mathscr{E}_{N}}(t)), t \geq 0,
		\end{equation*}
	\end{linenomath}
	and represents the index of the set $\mathscr{E}_{N}^{s}$ that the trace process $\xi_{N}^{\mathscr{E}_{N}}(\cdot)$ is at each time. 
	
	Denote by $D(\mathbb{R}_{+},\mathcal{H}_{N})$ the space of right-continuous functions $\boldsymbol{x}: \mathbb{R}_{+} \to \mathcal{H}_{N}$ with left-limits, endowed with the Skorohod topology and its associated Borel $\sigma$-field. Let $\boldsymbol{P}^{N}_{\eta}, \eta \in \mathcal{H}_{N}$, be the probability measure on $D(\mathbb{R}_{+},\mathcal{H}_{N})$ induced by the process $\xi_{N}(\cdot)$ starting from $\eta \in \mathcal{H}_{N}$. Expectation with respect to $\boldsymbol{P}_{\eta}^{N}$ is denoted by $\boldsymbol{E}_{\eta}^{N}$. Denote by $\mathbb{Q}_{\eta}^{N}, \eta \in \mathcal{H}_{N},$ the probability measure on $D(\mathbb{R}_{+},S)$ induced by measure $\boldsymbol{P}_{\eta}^{N}$ and the order process $Y_{N}(\cdot)$.
	
	\subsection{Metastability}
	
	Let $\mathcal{L}$ be the generator of a $S$-valued, continuous time Markov chain, given by
	\begin{linenomath}
		\begin{equation*}
			(\mathcal{L}f)(s) = \sum\limits_{r \in S} \mathfrak{r}(s,r) [f(r) - f(s)], s \in S,		
		\end{equation*}
	\end{linenomath}
	for $f: S \to \mathbb{R}$ and jump rates $\mathfrak{r}: S \times S \to \mathbb{R}_{+}$. The Markov process generated by $\mathcal{L}$ is such that a jump from $s$ to $r$ occurs at a rate $\mathfrak{r}(s,r)$. Denote by $\mathbb{Q}_{s}^{\mathcal{L}},s \in S,$ the probability measure on $D(\mathbb{R}_{+},S)$ induced by the Markov chain whose generator is $\mathcal{L}$ and which starts from $s$. The definition of metastability relies on two conditions.  The first one states that the measure $\mathbb{Q}_{\eta}^{N}$ converge to $\mathbb{Q}_{s}^{\mathcal{L}}$ when $\eta \in \mathscr{E}_{N}^{s}$ for all $N$. This implies that the limit of the order process is a Markov process with generator $\mathcal{L}$.
	
	\vspace{0.25cm}
	
	\noindent \textbf{Condition} $\mathfrak{C}_{\mathcal{L}}$. For all $s \in S$ and sequence $(\eta^{N})_{N \in \mathbb{Z}_{+}}$, such that $\eta^{N} \in \mathscr{E}_{N}^{s}$ for all $N \in \mathbb{Z}_{+}$, the sequence of laws $(\mathbb{Q}_{\eta^{N}}^{N})_{N \in \mathbb{Z}_{+}}$ converges to $\mathbb{Q}_{s}^{\mathcal{L}}$ as $N \to \infty$.
	
	\vspace{0.25cm}
	
	The second condition states that the process $\xi_{N}(\cdot)$ spends a negligible amount of time in $\Delta_{N}$ on each finite time interval when starting from a point in a well.
	
	\vspace{0.25cm}
	
	\noindent \textbf{Condition} $\mathfrak{D}$. For all $t > 0$,
	\begin{linenomath}
		\begin{equation*}
			\lim\limits_{N \to \infty} \max_{s \in S} \sup\limits_{\eta \in \mathscr{E}_{N}^{s}} \boldsymbol{E}_{\eta}^{N} \left[\int_{0}^{t} \chi_{\Delta_{N}}(\xi_{N}(s)) \ ds\right] = 0.
		\end{equation*}
	\end{linenomath}
	
	\vspace{0.25cm}
	
	A process is metastable if both conditions are satisfied. This definition is due to \cite{beltran2010tunneling} and is formalized below.
	
	\begin{definition}
		The process $\xi_{N}(\cdot)$ is said to be $\mathcal{L}$-metastable if conditions $\mathfrak{C}_{\mathcal{L}}$ and $\mathfrak{D}$ hold.
	\end{definition}
	
	Informally, the process $\xi_{N}(\cdot)$ is metastable when it spends a negligible time outside the wells $\mathscr{E}_{N}$ (condition $\mathfrak{D}$) and the dynamic determining its visits to the wells can be approximated by a simple Markov chain in $S$ (condition $\mathfrak{C}_{\mathcal{L}}$), when $N$ increases.
	
	\section{Hidden Markov processes and metastability}
	\label{sec_hidden_meta}
	
	In this section, we will formally define a model compatible with the assumptions in Section \ref{sec_theoreticalModel} and restate Theorem \ref{theorem_intro} as Theorem \ref{main_theorem}. In what follows, we assume $\xi_{N}(\cdot)$ is already speeded-up, so it is on the same timescale as the process $X_{N}(\cdot)$, and is $\mathcal{L}$-metastable, so assumptions \textbf{A1}, \textbf{A2} and \textbf{A3} are in force. We assume that the state space $\mathcal{X}$ of $X_{N}(\cdot)$ is countable to easy the presentation, but other cases could be considered.
	
	Given a countable set $\mathcal{X}$, consider the irreducible, continuous-time Markov process $Z_{N}(t)$$\coloneqq (\xi_{N}(t),X_{N}(t))$ taking values in $\mathcal{Z}_{N} \coloneqq \mathcal{H}_{N} \times \mathcal{X}$ with generator $\mathcal{L}_{Z_{N}}$ that acts on functions $\mathscr{F}_{N}: \mathcal{Z}_{N} \to \mathbb{R}$ as
	\begin{linenomath}
		\begin{equation}
			\label{generatorZN}
			(\mathcal{L}_{Z_{N}}\mathscr{F})(\eta,x) = (\mathcal{L}_{N}\mathscr{F}_{N,x})(\eta) + (\mathcal{L}_{X_{N}}\mathscr{F}_{N})(\eta,x),
		\end{equation}
	\end{linenomath}
	for $(\eta,x) \in \mathcal{Z}$, in which $\mathscr{F}_{N,x}: \mathcal{H}_{N} \to \mathbb{R}$ is defined as $\mathscr{F}_{N,x}(\eta) = \mathscr{F}_{N}(\eta,x)$ and
	\begin{linenomath}
		\begin{equation}
			\label{generatorXN}
			(\mathcal{L}_{X_{N}}\mathscr{F}_{N})(\eta,x) = \sum_{y \in \mathcal{X}} \gamma_{N}(\eta,y)\left[\mathscr{F}_{N}(\eta,y) - \mathscr{F}_{N}(\eta,x)\right]
		\end{equation}
	\end{linenomath} 
	in which $\gamma_{N}(\eta,y)$ is the jumping rate from $(\eta,x)$ to $(\eta,y)$. We assume that
	\begin{linenomath}
		\begin{equation}
			\label{limzero}
			\lim\limits_{N \to \infty} \gamma_{N}(\eta^{N},y) = 0
		\end{equation}
	\end{linenomath}
	for all $y \in \mathcal{X}$ if $\eta^{N} \notin \mathscr{E}_{N}$ for all $N \geq 1$. Recall that generator $\mathcal{L}_{N}$, defined in \eqref{generatorLN}, is that of $\xi_{N}(\cdot)$.
	
	The jumps of process $Z_{N}(\cdot)$ occur one coordinate at a time, that is from $(\eta,x)$ to a $(\xi,x)$ or to a $(\eta,y)$, while jumping to a $(\xi,y), x \neq y, \eta \neq \xi,$ is forbidden. The jump from $(\eta,x)$ to $(\xi,x)$ occurs as in process $\xi_{N}(\cdot)$, while the jump from $(\eta,x)$ to $(\eta,y)$ occurs at a rate that depends solely on $\eta$ and $y$, and not on $x$. While $Z_{N}(\cdot)$ and $\xi_{N}(\cdot)$ are Markov processes themselves, the process $X_{N}(\cdot)$ is not Markov. However, the process $X_{N}(\cdot)$ is dictated by the hidden Markov process $\xi_{N}(\cdot)$ and $Z_{N}(\cdot)$ is a HMM. Observe that these conditions are the first part of assumption \textbf{A4}. 
	
	We assume there exists a function $\gamma: S \times \mathcal{X} \to \mathbb{R}$ such that, for all $s \in S$,
	\begin{linenomath}
		\begin{equation}
			\label{assum_prob}
			\lim\limits_{N \to \infty} \sup\limits_{\eta \in \mathscr{E}_{N}^{s}} \sup\limits_{y \in \mathcal{X}} \left|\gamma_{N}(\eta,y) - \gamma(s,y)\right| = 0,
		\end{equation}
	\end{linenomath}
	which implies that, when $N$ tends to infinity, the jump rates from $(\xi,x)$ to $(\xi,y)$ are constant inside each metastable well $\mathscr{E}_{N}^{s}$ of $\xi_{N}(\cdot)$. That is the second part of assumption \textbf{A4}, so all four assumptions are in force in this model.
	
	Let $\mathcal{L}_{X}$ be the generator that acts on functions $\mathfrak{f}: S \times \mathcal{X} \to \mathbb{R}$ as
	\begin{linenomath}
		\begin{equation}
			\label{generatorLX}
			(\mathcal{L}_{X}\mathfrak{f})(s,x) = \sum_{y \in \mathcal{X}} \gamma(s,y) \left[\mathfrak{f}(s,y) - \mathfrak{f}(s,x)\right],
		\end{equation}
	\end{linenomath}
	and let $(Z(t) = (Y(t),X(t)): t \geq 0)$ be the process with state space $\mathcal{Z} = S \times \mathcal{X}$ and generator $\mathcal{L}_{Z}$ that acts on functions $\mathfrak{f}: S \times \mathcal{X} \to \mathbb{R}$ as
	\begin{linenomath}
		\begin{equation}
			\label{generatorZ}
			(\mathcal{L}_{Z}\mathfrak{f})(s,x) = (\mathcal{L}\mathfrak{f}_{x})(s) + (\mathcal{L}_{X}\mathfrak{f})(s,x),
		\end{equation}
	\end{linenomath}
	in which $\mathfrak{f}_{x}: S \to \mathbb{R}$ is defined as $\mathfrak{f}_{x}(s) = \mathfrak{f}(s,x)$ for $(s,x) \in \mathcal{Z}$. Recall that $\xi_{N}(\cdot)$ is $\mathcal{L}$-metastable and, since $Y(\cdot)$ is the process with generator $\mathcal{L}$, it is the limit of the order process. The generator in \eqref{generatorZ} is that of a hidden Markov model.
	
	Consider the wells
	\begin{linenomath}
		\begin{align}
			\label{wellsZ}
			\mathscr{Z}_{N}^{s,x} \coloneqq \mathscr{E}_{N}^{s} \times \{x\} & & \mathscr{Z}_{N} = \bigcup\limits_{\substack{s \in S\\ x \in \mathcal{X}}} \mathscr{Z}_{N}^{s,x}
		\end{align}
	\end{linenomath}
	with $\Delta_{N}^{Z} = \mathcal{Z}_{N} \setminus \mathscr{Z}_{N}$ and the projection $\Psi_{N}^{Z}: \mathscr{Z}_{N} \to S \times \mathcal{X}$ given by 
	\begin{linenomath}
		\begin{equation}
			\label{projectionZ}
			\Psi_{N}^{Z}(\eta,x) = \left(\sum_{s \in S} s \chi_{\mathscr{E}_{N}^{s}}(\eta),x\right).
		\end{equation}
	\end{linenomath}
	The main result of this paper is that $Z_{N}(\cdot)$ is $\mathcal{L}_{Z}$-metastable as stated in Theorem \ref{main_theorem}, which is the formalization of Theorem \ref{theorem_intro}. The proof of Theorem \ref{main_theorem} is in Appendix \ref{sec_proof}.
	
	\begin{theorem}
		\label{main_theorem}
		If $\xi_{N}(\cdot)$ is $\mathcal{L}$-metastable, then $Z_{N}(\cdot)$ is $\mathcal{L}_{Z}$-metastable with wells defined in \eqref{wellsZ} and projection defined in \eqref{projectionZ}. In special, the process $Z_{N}(\cdot) = (Y_{N}(\cdot),X_{N}(\cdot))$ converges to the hidden Markov model $Z(\cdot) = (Y(\cdot),X(\cdot))$.
	\end{theorem}
	
	\begin{remark}
		Since $\xi_{N}(\cdot)$ is $\mathcal{L}$-metastable, it spends a negligible amount of time not in $\mathscr{E}_{N}$ (cf. condition $\mathfrak{D}$) when starting from a well, as $N$ increases. Therefore, there is \textit{not enough time} for a jump from $(\eta,x)$ to $(\eta,y)$ if $\eta \notin \mathscr{E}_{N}$ when $N$ goes to infinity, and hence there is no loss of generality to assume that $\lim\limits_{N \to \infty} \gamma_{N}(\eta^{N},y) = 0$ if $\eta^{N} \notin \mathscr{E}_{N}$ for all $N \geq 1$. This assumption simplifies the proof of Theorem \ref{main_theorem}.
	\end{remark}
	
	\section{Example: Evolution of an asset price}
	\label{sec_example}
	
	In this section, we present an example of an interacting Markov system for modeling financial markets, that is adapted from an interacting particles system of physics.
	
	Assume the $N$ agents in a market $\mathcal{M}_{N}$ are, at each time $t \geq 0$, divided into two disjoint groups, denoted $G_{+1}$ and $G_{-1}$. The agents in $G_{+1}$ have the \textit{opinion}, or the \textit{will}, that the price of an asset will go up, while the agents in $G_{-1}$ have the \textit{opinion}, or the \textit{will}, that the price of the asset will go down. In this instance, \textit{going up} means attaining a value $A$ and going down means attaining a value $B$, considering that the current value of the asset is greater than $B$ and lesser than $A$. 
	
	Assuming that the agents in each group are taking actions to \textit{enforce} their will, or to \textit{prepare} for the realization of their opinion, it is expected that the number of agents in each group will influence the price of the asset. Hence, the transition probabilities of the asset price can be modeled as a function of the proportion of agents in each group, regardless of its current and past prices. Proceeding in this way, the Markov process describing the market opinion can be a hidden dynamic that determines the price evolution of the asset, that is not itself Markov.
	
	In Section \ref{SecZRP} we present an example of an adapted interacting particles system in this context, in which its physical meaning is reinterpreted as a financial market. In Section \ref{SecSimulation} we present a simulation of this model to illustrate the main ideas discussed in this paper.
	
	\subsection{Zero-range process}
	\label{SecZRP}
	
\begin{figure}[ht]
		\centering
		\begin{tikzpicture}[every node/.style={inner sep=0,outer sep=0}]
			\tikzstyle{hs} = [circle,fill,draw=black, rounded corners,minimum width=0.8em, node distance=2.5cm, line width=1pt]
			
			\node[hs] at (1,0) {};
			\node[hs] at (1,0.28) {};
			\node[hs] at (1,0.28*2) {};
			\node[hs] at (1,0.28*3) {};
			\node[hs] at (1,0.28*4) {};
			\node[hs] at (1,0.28*5) {};
			\node[hs] at (1,0.28*6) {};
			\node[hs] at (1,0.28*7) {};
			\node[hs] at (1,0.28*8) {};
			\node[hs] at (1,0.28*9) {};
			\node[hs] at (1,0.28*10) {};
			\node[hs] at (1,0.28*11) {};
			\node[hs] at (1,0.28*12) {};
			\node[hs] at (1,0.28*13) {};
			\node[hs] at (1,0.28*14) {};
			\node[hs] at (1,0.28*15) {};
			\node[hs] at (1,0.28*16) {};
			\node[hs] at (1,0.28*17) {};
			\node[hs] at (1,0.28*18) {};
			\node[hs] at (1,0.28*19) {};
			\node[hs] at (1,0.28*20) {};
			
			\node[hs] at (5,0) {};
			\node[hs] at (5,0.28) {};
			\node[hs] at (5,0.28*2) {};
			\node[hs] at (5,0.28*3) {};
			\node[hs] at (5,0.28*4) {};
			
			\node at (1,-0.5) {$-1$};
			\node at (5,-0.5) {$1$};

			\begin{scope}[line width=1pt]
				\draw[-] (-0.25,-0.25) -- (6,-0.25);
				\draw[-] (0,-0.5) -- (0,0.28*23);
				\draw[-] (1,-0.35) -- (1,-0.15);
				\draw[-] (5,-0.35) -- (5,-0.15);
				
				\draw[->] (5,0.28*5) to[bend right] node[sloped,below] {$g(\eta_{+1})$} (1.25,0.28*21);
				\draw[->] (1.25,0.28*20) to[bend right] node[sloped,below] {$g(\eta_{-1})$} (4.75,0.28*5);
			\end{scope}
		\end{tikzpicture}
		\caption{\footnotesize Illustration of the zero-range process, in which an agent leaves the state $s \in \{-1,+1\}$ at a rate $g(\eta_{s})$ (cf. \eqref{def_g}) which is decreasing on the number $\eta_{s}$ of agents in $s$.} \label{fig_ZRP}
	\end{figure}
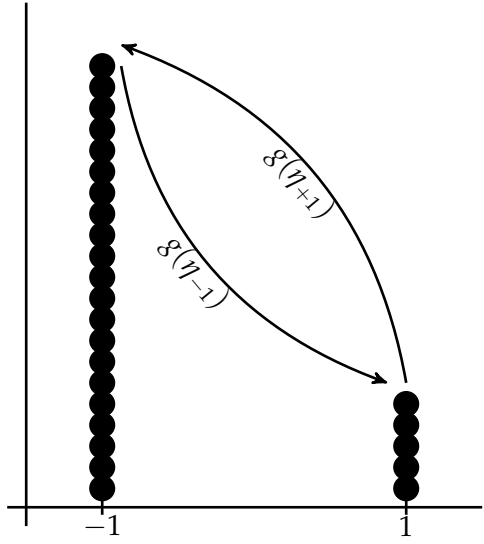

	Assume the agents change groups in \textit{real-time}, and the rate at which agents leave a group is decreasing on the number of agents in the group, as illustrated in Figure \ref{fig_ZRP}. This means that, as more agents have a same dominant opinion, lesser is the probability of an agent changing its opinion against the majority, and higher is the probability of a dissident agent changing its opinion to that of the majority.
	
	In such a dynamic, it is expected that a consensus opinion will prevail for a long time, until a growing number of dissidents cause the consensus to flip. Considering that the number of agents in each group evolves as a Markov process, this dynamic points to a metastable behavior, if the number of agents increases and the process is observed on the timescale of days, for example, instead of in \textit{real-time}. This is a reinterpretation of the zero-range process \cite{andjel1982invariant}, which is as follows.
	
	Let $S = \{-1,+1\}$ represents the two possible states of each agent in the market. A configuration $\eta \in \mathbb{Z}_{+}^{S}$ of agents is denoted by $\eta = (\eta_{-1},\eta_{+1})$, where $\eta_{-1},\eta_{+1}$ represent the number of agents at the respective state under configuration $\eta$. For $N \in \mathbb{Z}_{+}$, denote by $\mathcal{H}_{N} \subset \mathbb{Z}_{+}^{S}$ the subset of configurations with exactly $N$ agents:
	\begin{linenomath}
		\begin{equation*}
			\mathcal{H}_{N} = \left\{\eta = (\eta_{-1},\eta_{+1}) \in \mathbb{Z}_{+}^{S}: \eta_{-1} + \eta_{+1} = N\right\}.
		\end{equation*}
	\end{linenomath}
	The zero-range process is the continuous-time Markov chain $(\eta_{N}(t): t \geq 0)$ on $\mathcal{H}_{N}$ with generator acting on functions $F: \mathcal{H}_{N} \to \mathbb{R}$ as
	\begin{linenomath}
		\begin{equation*}
			(\tilde{\mathcal{L}}_{N}F)(\eta) = \sum_{s,r \in S} \mathfrak{r}(s,r) g(\eta_{s}) \left[F(\sigma^{s,r}\eta) - F(\eta)\right], \ \eta \in \mathcal{H}_{N}
		\end{equation*}
	\end{linenomath}
	in which $\mathfrak{r}: S^{2} \to \mathbb{R}_{+}$ is symmetric, $\mathfrak{r}(s,s) = 0,s \in S,$ and, for a fixed parameter $\alpha > 1$,
	\begin{linenomath}
		\begin{align}
			\label{def_g}
			g(0) = 0, & & g(1) = 1, & & \text{ and } & & g(n) = \left(\frac{n}{n-1}\right)^{\alpha} \text{ for } n \geq 2.
		\end{align}
	\end{linenomath}
	The notation $\sigma^{s,r}\eta, s,r \in \{-1,+1\}$, stands for the configuration obtained from $\eta$ by moving an agent from $s$ to $r$, when there is at least one agent at $s$:
	\begin{linenomath}
		\begin{align*}
			(\sigma^{s,r}\eta)_{u} = \begin{cases}
				\eta_{s} - 1 & \text{ if } u = s,\\
				\eta_{r} + 1 & \text{ if } u = r,
			\end{cases} & & u \in \{-1,+1\}
		\end{align*}
	\end{linenomath}
	if $\eta_{s} \geq 1$. Otherwise, $\sigma^{s,r}\eta = \eta$. The process generated by $\tilde{\mathcal{L}}_{N}$ jumps from $\eta = (\eta_{-1},\eta_{+1})$ to $(\eta_{-1} - 1,\eta_{+1} + 1)$ with rate $\mathfrak{r}(-1,1) g(\eta_{-1})$ and to $(\eta_{-1} + 1,\eta_{+1} - 1)$ with rate $\mathfrak{r}(1,-1) g(\eta_{+1})$.
	
	Denote $\theta_{N} = N^{1+\alpha}$ for $N \in \mathbb{Z}_{+}$, and consider the speeded-up process $(\xi_{N}(t): t \geq 0)$ defined as $\xi_{N}(t) = \eta_{N}(\theta_{N}t)$. The process $\xi_{N}(\cdot)$ is the continuous-time Markov process on $\mathcal{H}_{N}$ with generator
	\begin{linenomath}
		\begin{equation*}
			(\mathcal{L}_{N}F)(\eta) = (\theta_{N}\tilde{\mathcal{L}}_{N}F)(\eta) = \sum\limits_{\xi \in \mathcal{H}_{N}} \mathfrak{R}_{N}(\eta,\xi) \left[F(\xi) - F(\eta)\right], \eta \in \mathcal{H}_{N},
		\end{equation*}
	\end{linenomath}
	in which
	\begin{linenomath}
		\begin{equation*}
			\mathfrak{R}_{N}(\eta,\xi) = \begin{cases}
				\theta_{N} \mathfrak{r}(s,r) g(\eta_{s}), & \text{ if } \xi = \sigma^{s,r}\eta \text{ for } s,r \in S\\
				0, & \text{ otherwise}.
			\end{cases}
		\end{equation*}
	\end{linenomath}
	
	Let
	\begin{linenomath}
		\begin{align*}
			\mathscr{E}_{N}^{-1} = \{\eta \in \mathcal{H}_{N}: \eta_{-1} \geq N - \ell_{N}\} & & \mathscr{E}_{N}^{+1} = \{\eta \in \mathcal{H}_{N}: \eta_{+1} \geq N - \ell_{N}\}
		\end{align*}
	\end{linenomath}
	be the metastable wells of $\xi_{N}(\cdot)$, in which $\ell_{N}$ is any sequence satisfying 
	\begin{linenomath}
		\begin{align*}
			\lim\limits_{N \to \infty} \ell_{N} =  \infty & & \text{ and } & &  \lim\limits_{N \to \infty} \frac{\ell_{N}}{N} = 0.
		\end{align*}
	\end{linenomath}
	The sets $\mathscr{E}_{N}^{-1}$ and $\mathscr{E}_{N}^{+1}$ contain the configurations on which a consensus has been reached.
	
	Consider the generator $\mathcal{L}$ of a Markov chain in $S$ that acts on functions $f: S \to \mathbb{R}$ as
	\begin{linenomath}
		\begin{equation}
			\label{generator_microZRP}
			(\mathcal{L}f)(s) = \frac{1}{\Gamma(\alpha) I_{\alpha}} \sum_{r \in S} cap(r,s)[f(r) - f(s)].
		\end{equation}
	\end{linenomath}	
	in which $\Gamma(\alpha)$ is the gamma function, $I_{\alpha} = \int_{0}^{1} u^{\alpha} (1-u)^{\alpha} \ du$ and $cap(r,s)$ is the capacity between sets $\{r\}$ and $\{s\}$ under the Markov chain with rates $\mathfrak{r}(\cdot,\cdot)$ (see \cite[(2.3)]{beltran2012metastability} for a definition of the capacity). It was shown in \cite{beltran2012metastability} that the process $\xi_{N}(\cdot)$ is $\mathcal{L}$-metastable.
	
	\subsection{Causal model for an asset price evolution}
	
	From the main result of this paper follows that, by properly accelerating the zero-range process described above, and making the number of agents increase, the dynamics of the two-dimensional process whose coordinates represent the consensus opinion and the asset price variation will be approximated by a HMM. We now formally define this model.
	
	Consider the Markov process $(Z_{N}(t) = (\xi_{N}(t),X_{N}(t)): t \geq 0)$ on $\mathcal{Z}_{N} = \mathcal{H}_{N} \times \{-1,+1\}$ with generator acting on functions $\mathscr{F}: \mathcal{Z}_{N} \to \mathbb{R}$ as
	\begin{linenomath}
		\begin{equation}
			\label{gen_example}
			(\mathcal{L}_{Z_{N}}\mathscr{F})(\eta,x) = (\mathcal{L}_{N}\mathscr{F}_{N,x})(\eta) + (\mathcal{L}_{X_{N}}\mathscr{F}_{N})(\eta,x)
		\end{equation}
	\end{linenomath}
	for $(\eta,x) \in \mathcal{Z}$, in which $\mathscr{F}_{N,x}: \mathcal{H}_{N} \to \mathbb{R}$ is defined as $\mathscr{F}_{N,x}(\eta) = \mathscr{F}_{N}(\eta,x)$ and
	\begin{linenomath}
		\begin{equation*}
			(\mathcal{L}_{X_{N}}\mathscr{F}_{N})(\eta,x) = \sum_{y \in \{-1,+1\}} \gamma_{N}(\eta,y)\left[\mathscr{F}_{N}(\eta,y) - \mathscr{F}_{N}(\eta,x)\right]
		\end{equation*}
	\end{linenomath} 
	where $\gamma_{N}: \mathcal{Z}_{N} \to \mathbb{R}$ is defined as $\gamma_{N}(\eta,-1) = \delta(1 - P_{N}(\eta))$ and $\gamma_{N}(\eta,+1) = \delta P_{N}(\eta)$ for a $\delta > 0$ and $\eta \in \mathcal{H}_{N}$, with $P_{N}: \mathcal{H}_{N} \to [0,1]$ given by
	\begin{linenomath}
		\begin{equation*}
			P_{N}(\eta) = \frac{e^{a + b \frac{\eta_{+1}}{N}}}{1 + e^{a + b \frac{\eta_{+1}}{N}}} \chi_{\mathscr{E}_{N}}(\eta)
		\end{equation*}
	\end{linenomath}
	with $a \leq 0$ and $b \geq 0$. The probability $P_{N}(\eta)$ is that of $Z_{N}(\cdot)$ jumping to $(\eta,+1)$ conditioned on it jumping to either $(\eta,+1)$ or $(\eta,-1)$.
	
	Observe that
	\begin{linenomath}
		\begin{equation*}
			\lim\limits_{N \to \infty} P_{N}(\eta^{N}) = \lim\limits_{N \to \infty} \frac{e^{a + b \frac{\eta_{+1}^{N}}{N}}}{1 + e^{a + b \frac{\eta_{+1}^{N}}{N}}} = \begin{cases}
				\frac{e^{a + b}}{1 + e^{a + b}}, & \text{ if } \eta^{N} \in \mathscr{E}_{N}^{+1}, \forall N \in \mathbb{Z}_{+},\\
				\frac{e^{a}}{1 + e^{a}}, & \text{ if } \eta^{N} \in \mathscr{E}_{N}^{-1}, \forall N \in \mathbb{Z}_{+},
			\end{cases}
		\end{equation*}
	\end{linenomath}
	and hence $\gamma_{N}(\eta,y)$ converges to a $\gamma(s,y)$  for all $\eta \in \mathscr{E}_{N}^{s}$ in which $\gamma(s,-1) = \delta(1 - P(s))$ and $\gamma(s,+1) = \delta P(s)$, with
	\begin{linenomath}
		\begin{align*}
			P(+1) = \frac{e^{a + b}}{1 + e^{a + b}} & & \text{ and } & & P(-1) = \frac{e^{a}}{1 + e^{a}}.
		\end{align*}
	\end{linenomath}
	The probabilities $P(+1)$ and $P(-1)$ will be the transition probabilities to $+1$ of the observed chain $X(\cdot)$ when the hidden state is $+1$ and $-1$, respectively.
	
	It follows from Theorem \ref{main_theorem} that the process $Z_{N}(\cdot)$ is $\mathcal{L}_{Z}$-metastable with 
	\begin{linenomath}
		\begin{equation*}
			(\mathcal{L}_{Z}\mathfrak{f})(s,x) = (\mathcal{L}\mathfrak{f}_{x})(s) + (\mathcal{L}_{X}\mathfrak{f})(s,x)
		\end{equation*}
	\end{linenomath}
	in which $\mathcal{L}$ is defined in \eqref{generator_microZRP} and
	\begin{linenomath}
		\begin{equation*}
			(\mathcal{L}_{X}\mathfrak{f})(s,x) = \sum_{y \in \{-1,+1\}} \gamma(s,y) \left[\mathfrak{f}(s,y) - \mathfrak{f}(s,x)\right].
		\end{equation*}
	\end{linenomath}
	We denote the Markov process with generator $\mathcal{L}_{Z}$ by $Z(t) = (Y(t),X(t))$ in which $Y(\cdot)$ is the process with generator $\mathcal{L}$ and $X(\cdot)$ is the limit of the process $X_{N}(\cdot)$.
	
	Denote $t_{1}, t_{2}, t_{3}, \dots$ the jumping times\footnote{Observe that process $X(\cdot)$ can ``jump'' to the same state it jumped to lastly, .i.e., it has a probability of staying at the same state.} of process $X(\cdot)$. Under this model the asset price at time $t \geq 0$ is given by the process
	\begin{linenomath}
		\begin{equation*}
			S(t) = S_{0} + \sum_{n = 1}^{\infty} X(t_{n}) \ \mathds{1}\{t_{n} \leq t\}
		\end{equation*}
	\end{linenomath}
	in which $S_{0}$ is the asset price at $t = 0$. Therefore, the process $X(\cdot)$ represents the variation of the asset price: when $X(\cdot)$ jumps to $+1$ the asset price increases in one unit and when it jumps to $-1$ it decreases in one unit. Hence, variations on the asset price are caused by the market agents through the causal mechanism represented by the generator $\mathcal{L}_{Z}$.
	
	The dynamics described by $\mathcal{L}_{Z}$ is as follows. Denote by $\{X_{n}: n \in \mathbb{Z}_{+}\}$ the process defined as $X_{n} = X(t_{n})$, that is not Markov. We have that, conditioned on $Y(t_{n})$, the transition probabilities of $X_{n}$ are the following:
	\begin{linenomath}
		\begin{equation*}
			\mathbb{P}\left(X_{n} = 1|Y(t_{n})\right) = 1 - \mathbb{P}\left(X_{n} = -1|Y(t_{n})\right) = P(Y(t_{n})).
		\end{equation*}
	\end{linenomath}
	This implies that, when the consensus is that the asset price will go up, then the probability of it going up is $P(+1) = \frac{e^{a + b}}{1 + e^{a + b}}$, while when the consensus is that the asset price is going down, then the probability of it going up is $P(-1) = \frac{e^{a}}{1 + e^{a}}$.
	
	The values of $a$ and $b$ is a measure of how strong is the impact of a consensus on the asset price. For example, if $b \gg |a|$ then $\frac{e^{a + b}}{1 + e^{a + b}} \approx 1$, so a consensus on the growth of the price almost causes it to go up. In the same manner, if $|a| \gg 0$, then $\frac{e^{a}}{1 + e^{a}} \approx 0$, so a consensus about the price going down practically causes it to go down. If $b \approx 0$, then these probabilities are equal and the opinion of the market does not interfere on the asset price. Also, if $\frac{e^{a + b}}{1 + e^{a + b}} > 1/2$ then a consensus on the growth of the price causes it to be more likely to happen than not, and if $\frac{e^{a}}{1 + e^{a}} < 1/2$ then a consensus about the price going down causes it to be more likely to happen than not.
	
	This model is a simplification of a real market and the influence of its agents on a asset price. Nevertheless, it illustrates how the hypothesis that a hidden Markov chain dictates the evolution of an asset price is a consequence of a theory that describes the interaction between market agents as metastable interacting Markov systems.
	
	\begin{remark}
		We assumed the rates $\mathfrak{r}(s,r)$ are symmetric to simplify the generator $\mathcal{L}$, but the result also holds when they are not symmetric \cite{beltran2012metastability}. If the agents are assumed to be divided in more than two groups, and the number of agents in each group evolves as a zero-range process as that described in \cite{landim2023metastable,landim2021resolvent} for $\alpha = 1$ (critical) or $\alpha > 1$ \cite{beltran2012metastability} (supercritical), then a metastable behavior may be observed if $P_{N}$ is suitably chosen. We considered only two sites in the zero-range process, that is actually a random walk in $\{1,\dots,N\}$, in order to illustrate the main ideas of the approach and its consequences, and more sites could be considered at the cost of heavier notation.
	\end{remark}
	
	\subsection{Model simulation}
	\label{SecSimulation}
	
	In order to illustrate the main ideas discussed in this paper, we present a simulation of the model for the evolution of an asset price presented above. We considered $N = 10,000$ and the other model parameters as described in Table \ref{tabPar}. In special, we considered the wells $\mathscr{E}_{N}^{-1}$ and $\mathscr{E}_{N}^{+1}$ as containing the configurations with at least 2/3 of the agents in $G_{-1}$ and $G_{+1}$, respectively, so the configurations in $\Delta_{N}$ are those in which both $G_{-1}$ and $G_{+}$ have at least 1/3 of the market agents. Finally, we chose $a$ and $b$ such that\footnote{We considered $P_{N}(\eta) = \frac{e^{a + b \frac{\eta_{+1}}{N}}}{1 + e^{a + b \frac{\eta_{+1}}{N}}}$, so it is not zero in $\Delta_{N}$. Since $N$ is finite, jumps of $X_{N}(\cdot)$ can happen while $\xi_{N}(\cdot)$ is in $\Delta_{N}$ so defining $P_{N}$ in this is way makes more sense.}
	\begin{linenomath}
		\begin{align*}
			P_{N}(\eta) \approx 0.525 \text{ for } \eta \in \mathscr{E}_{N}^{+1} & & \text{ and } & &  P_{N}(\eta) \approx 0.475 \text{ for } \eta \in \mathscr{E}_{N}^{-1}.
		\end{align*}
	\end{linenomath}
	Under this choice of parameters, the process $X_{N}(\cdot)$ is assumed to be in the scale of days, so the market process is in the scale of an order $\theta^{-1} \approx 10^{-8}$ of a day. More precisely, the market process is in the scale of 0.78 ms. As a reference, the time it takes to blink is around 100 ms, so while one blinks, he can expect dozens of jumps to have occurred on the market process.
	
	\begin{table}[ht]
		\centering
		\caption{\footnotesize Parameters of the simulated model.} \label{tabPar}
		\begin{tabular}{cc}
			\hline
			Parameter & Value \\
			\hline 
			N & 10,000\\
			$\alpha$ & 1.01\\
			$\theta$ & 109,647,820 \\
			$\delta$ & 5 \\
			$\mathfrak{r}(s,r)$ & 0.1 \\
			a & -0.1000835 \\
			b & 0.2001669 \\
			$\ell_{N}$ & 3,333 \\
			$S_{0}$ & 100 \\
			\hline
		\end{tabular}
	\end{table}
	
	\begin{figure}[ht]
		\centering
		\includegraphics[scale = 0.45]{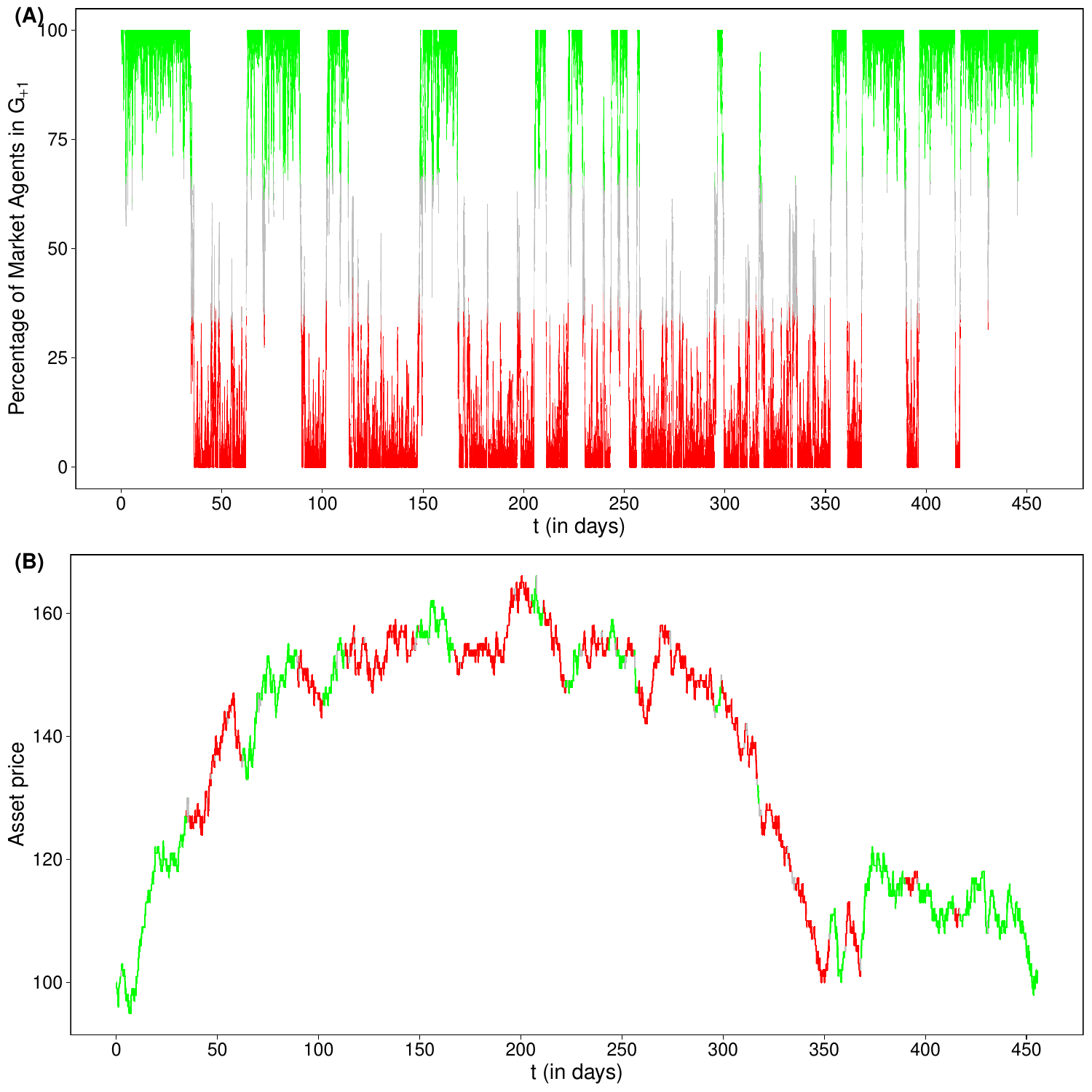}
		\caption{\footnotesize A simulation of the the process with generator $\mathcal{L}_{Z_{N}}$ defined in \eqref{gen_example} with the parameters in Table \ref{tabPar}. \textbf{(A)} The percentage of the market agents in $G_{+1}$ at each time $t$. \textbf{(B)} The asset price in the time-scale of days at each time $t$. The colors represents the times that the process $\xi(\cdot)$ is in the wells $\mathscr{E}_{N}^{+1}$ (green) and $\mathscr{E}_{N}^{-1}$ (red), and in the set $\Delta_{N}$ (gray).} \label{fig_simulation}
	\end{figure}
	
	Starting from all agents in $G_{+1}$, we simulated $10^{10}$ jumps of the zero-range process. It took 455 days for these jumps to happen, so in average the market changed its state 21,978,022 times in a day, or around 25 times every 100 ms. In Figure \ref{fig_simulation} we present the percentage of the market agents in $G_{+1}$ and the asset price at each time. We use colors to outline the hidden states of the market, characterized when the process is in $\mathscr{E}_{N}^{+1}$ (green) and $\mathscr{E}_{N}^{-1}$ (red). We also consider a color when the process is in $\Delta_{N}$ (gray). A video portraying the evolution of the market state and the asset price is available at \url{https://www.youtube.com/watch?v=bwSQTxba3DA} and we encourage the reader to watch it to see how the evolution unfolds.
	
	In Figure \ref{fig_simulation} \textbf{(A)} we observe the metastable behavior of the market. When the process reaches a well, it spends an exponential time in it, that can last up to two months. There are some very quick excursions of the process to $\Delta_{N}$, after which it returns to the well it departed from. The transition between wells occurs abruptly, so the process does not spend time in $\Delta_{N}$. Furthermore, once the process reaches the other well, it attains abruptly its bottom, i.e., the configuration with all agents in a same group, and the same behavior starts again, now from the other well.
	
	In Figure \ref{fig_simulation} \textbf{(B)} we can see the effect of the market state on the asset price. When the hidden state of the market is $+1$ (green) there is an upward trend in the asset price, and when the hidden state is $-1$ (red) there is a downward trend. This is the case since $P_{N}(\eta) > 1/2$ for $\eta \in \mathscr{E}_{N}^{+1}$ and $P_{N}(\eta) < 1/2$ for $\eta \in \mathscr{E}_{N}^{-1}$. The causal relation between the asset price and the market state is clear when we see periods of sudden persistence of growth and decline that are due to changes of the hidden state. Overall, the plot in Figure \ref{fig_simulation} \textbf{(B)} has the characteristics one usually observes in financial time-series and a causal relation between its variation and the market state can be identified.

	\section{Discussion}
	\label{sec_discuss}
	
	Hidden Markov models have been applied to model financial time series for decades and the main contribution of this paper is the deduction of a causal mechanism that justifies such modeling. We showed that, under plausible assumptions, variations in a financial time series are caused by a metastable interacting Markov system representing a financial market. The theory we proposed is a rare application of metastable Markov processes outside the modeling of physical systems.\footnote{A rare application of metastability apart from random perturbations of dynamical systems, which can be applied in many domains.}
	
	The proposed theory opens up the perspective of studying new metastable interacting Markov systems inspired by financial markets. The interacting particle systems studied in the literature are usually constrained by physical laws, and many models with alternative constraints have yet to be explored. If an interacting Markov system is though of as a financial market, then other constraints arise and a new family of systems can be proposed and their properties studied. The development of such systems is a topic for future research.
	
	From the perspective of factor investing, scientific practical methods may be developed to test the fitness of market interacting Markov systems and of the modeling by the implied hidden Markov model. Recalling that market agents are not only \textit{people}, but also algorithms, we argue that the state of the market can in principle be assessed by pooling algorithms settings, and the opinion and strategies of other agents, and based on them the hidden state can be estimated and a prediction for a financial time series could be generated. These predictions could also be generated by classical methods for the inference of HMM \cite{zucchini2017hidden}. If the predictions are not substantiated by empirical data, then the causal relation generated by the HMM is falsified, hence these methods are indeed scientific. The perspective of developing such practical methods is promising.
	
	The standing paradigm of causal factor investing, based on causal graphs and tools from causal statistical inference \cite{pearl2016causal}, is not completely inconsistent with the proposed theory. The observed time series can actually be multidimensional and, depending on how the market state influences the time series, there may exist a causal, mediation and/or moderation relation between coordinates (i.e., variables) of the time series. Hence, the theory proposed here can be combined with causal inference to model financial markets. 
	
	Nevertheless, the proposed theory has an advantage over causal graphs since accounts for the non-stationary evolution of the time series through the change between metastable wells. Many causal inference techniques assume that patterns observed in the past will repeat themselves in the future, what is not true in non-stationary evolutions. This is one of the features of financial time series that cause predictions based on causal inference to perform poorly out-of-sample in numerous instances. Accounting for the possibility of the market changing metastable wells, one can better assess the risks and, even if these changes cannot be predicted, they are expected and risk mitigating strategies may be put into place in case they happen. When one relies on an immutable causal graph, he is not accounting for the possibility of changes on it and hence is exposed to more risks than previously assessed.
	
	Metastability and HMM are classical topics in Markov processes that have been intensively studied for decades. The main contribution of this paper is to link these concepts in the context of finance, establishing a probabilistic theory for modeling financial markets.
	
	\bibliographystyle{plain}
	\bibliography{Ref}
	
	\appendix
	\section{Proof of Theorem \ref{main_theorem}}
	\label{sec_proof}
	
	The proof of Theorem \ref{main_theorem} relies on the resolvent approach to metastability. In the years following \cite{beltran2010tunneling}, many sufficient conditions for metastability have been deduced \cite{beltran2012tunneling,beltran2015martingale}. Although they were successfully applied to many processes \cite{landim2019metastable}, they rely, for instance, on explicit calculations of the stationary state and estimation of capacities, or on the fact that each well has an attractor. In \cite{landim2021resolvent} it was deduced a sufficient, and necessary, condition for metastability based on the solution of resolvent equations that does not necessarily depend on such calculations and may be deduced without relying on the existence of an attractor, or on other assumptions of earlier works. 
	
	The resolvent approach to metastability is as follows. Fix $g: S \to \mathbb{R}$, and let $G_{N}: \mathcal{H}_{N} \to \mathbb{R}$ be given by
	\begin{linenomath}
		\begin{equation*}
			G_{N}(\eta) = \sum_{s \in S} g(s)\chi_{\mathscr{E}_{N}^{s}}(\eta),
		\end{equation*}
	\end{linenomath}
	that is the function that equals $g(s)$ in $\mathscr{E}_{N}^{s}, s \in S,$ and zero in $\Delta_{N}$. For $\lambda > 0$, denote by $F_{N} = F_{N}^{\lambda,g}$ the unique solution of the resolvent equation
	\begin{linenomath}
		\begin{equation}
			\label{resolvent_eq}
			(\lambda - \mathcal{L}_{N}) F_{N} = G_{N}.
		\end{equation}
	\end{linenomath}
	In \cite{landim2021resolvent} it is shown that the following condition is equivalent to $\mathcal{L}$-metastability.
	
	\vspace{0.25cm}
	
	\noindent \textbf{Condition} $\mathfrak{R}_{\mathcal{L}}$. For all $\lambda > 0$ and $g: S \to \mathbb{R}$, the unique solution $F_{N}$ of the resolvent equation \eqref{resolvent_eq} is asymptotically constant in each set $\mathscr{E}_{N}^{s}$ and
	\begin{linenomath}
		\begin{equation*}
			\lim\limits_{N \to \infty} \sup\limits_{\eta \in \mathscr{E}_{N}^{s}} \left|F_{N}(\eta) - f(s)\right| = 0, s \in S,
		\end{equation*}
	\end{linenomath} 
	where $f: S \to \mathbb{R}$ is the unique solution of the reduced resolvent equation
	\begin{linenomath}
		\begin{equation*}
			(\lambda - \mathcal{L})f = g.
		\end{equation*}
	\end{linenomath}
	
	\vspace{0.25cm}
	
	The main result of \cite{landim2021resolvent} is the following.
	
	\begin{theorem}
		\label{theorem_resolvent}
		The process $\xi_{N}(\cdot)$ is $\mathcal{L}$-metastable if, and only if, condition $\mathfrak{R}_{\mathcal{L}}$ is satisfied.
	\end{theorem}
	
	We are now in position to prove Theorem \ref{main_theorem}.
	
	\begin{proof}
		The strategy to prove Theorem \ref{main_theorem} is to show that the generator $\mathcal{L}_{Z_{N}}$ satisfies condition $\mathfrak{R}_{\mathcal{L}_{Z}}$. This proof relies heavily on the unicity of resolvent equations to \textit{transfer} the metastability of $\xi_{N}(\cdot)$ to $Z_{N}(\cdot)$.	
		
		Fix $\lambda > 0$ and a function $\mathfrak{g}: S \times \mathcal{X} \to \mathbb{R}$, and let $\mathfrak{f}: S \times \mathcal{X} \to \mathbb{R}$ be the solution of the reduced resolvent equation
		\begin{linenomath}
			\begin{align}
				\label{resolvent_reduced_Z}
				\left[(\lambda - \mathcal{L}_{Z})\mathfrak{f}\right](s,x) = \mathfrak{g}(s,x) & & (s,x) \in S \times \mathcal{X}.
			\end{align}
		\end{linenomath}
		Let $\mathscr{G}_{N}: \mathcal{Z}_{N} \to \mathbb{R}$ be given by
		\begin{linenomath}
			\begin{equation*}
				\mathscr{G}_{N}(\eta,x) = \sum\limits_{s \in S} \mathfrak{g}(s,x) \chi_{\mathscr{E}_{N}^{s}}(\eta),
			\end{equation*}
		\end{linenomath}
		and denote by $\mathscr{F}_{N}: \mathcal{Z}_{N} \to \mathbb{R}$ the solution of resolvent equation
		\begin{linenomath}
			\begin{align}
				\label{resolventZ}
				\left[(\lambda - \mathcal{L}_{Z_{N}})\mathscr{F}_{N}\right](\eta,x) = \mathscr{G}_{N}(\eta,x) & & (\eta,x) \in \mathcal{Z}_{N}.
			\end{align}
		\end{linenomath}
		We need to show that
		\begin{linenomath}
			\begin{equation}
				\label{cond_to_show}
				\lim\limits_{N \to \infty} \sup\limits_{x \in \mathcal{X}} \sup\limits_{\eta \in \mathscr{E}_{N}^{s}} \left|\mathscr{F}_{N}(\eta,x) - \mathfrak{f}(s,x)\right| = 0, s \in S,
			\end{equation}
		\end{linenomath}
		so the $\mathcal{L}_{Z}$-metastability of $Z_{N}$ follows from Theorem \ref{theorem_resolvent}.
		
		Now, fix $x \in \mathcal{X}$ and denote $g_{x}(s) = \mathfrak{g}(s,x)$ for $s \in S$. Let $f_{x}: S \to \mathbb{R}$ be the solution of the reduced resolvent equation
		\begin{linenomath}
			\begin{align}
				\label{resolvent_reduced_Zmodified}
				\left[(\lambda - \mathcal{L})f_{x}\right](s) = g_{x}(s) + (\mathcal{L}_{X}\mathfrak{f})(s,x) & & s \in S.
			\end{align}
		\end{linenomath}
		Define
		\begin{linenomath}
			\begin{equation*}
				G_{N,x}(\eta) = \sum_{s \in S} \left[g_{x}(s) + (\mathcal{L}_{X}\mathfrak{f})(s,x)\right] \chi_{\mathscr{E}_{N}^{s}}(\eta)
			\end{equation*}
		\end{linenomath}
		for $\eta \in \mathcal{H}_{N}$ and let $F_{N,x}$ be the solution of resolvent equation
		\begin{linenomath}
			\begin{align}
				\label{resolvent_etaX}
				\left[(\lambda - \mathcal{L}_{N})F_{N,x}\right](\eta) = G_{N,x}(\eta), & & \eta \in \mathcal{H}_{N}.
			\end{align}
		\end{linenomath}
		Since $\xi_{N}(\cdot)$ is $\mathcal{L}$-metastable, it follows from Theorem \ref{theorem_resolvent} that, for every $x \in \mathcal{X}$ fixed,
		\begin{linenomath}
			\begin{equation}
				\label{implied}
				\lim\limits_{N \to \infty} \sup\limits_{\eta \in \mathscr{E}_{N}^{s}} \left|F_{N,x}(\eta) - f_{x}(s)\right| = 0, s \in S.
			\end{equation}
		\end{linenomath}
		
		In view of \eqref{implied}, in order to show \eqref{cond_to_show} it is enough to show that
		\begin{linenomath}
			\begin{equation}
				\label{equality_toShow}
				\mathfrak{f}(s,x) = f_{x}(s)
			\end{equation}
		\end{linenomath}
		for all $s \in S$ and all $x \in \mathcal{X}$, and that
		\begin{linenomath}
			\begin{equation}
				\label{limit_toShow}
				\lim\limits_{N \to \infty} \sup\limits_{x \in \mathcal{X}} \sup\limits_{\eta \in \mathscr{E}_{N}^{s}} \left|F_{N,x}(\eta) - \mathscr{F}_{N}(\eta,x)\right| = 0, s \in S.
			\end{equation}
		\end{linenomath}
		Indeed, \eqref{cond_to_show} follows by combining \eqref{implied}, \eqref{equality_toShow} and \eqref{limit_toShow}.
		
		We start by showing that \eqref{equality_toShow} holds. Define $h_{x}(s) \coloneqq \mathfrak{f}(s,x)$ for $s \in S, x \in \mathcal{X}$. Observe that
		\begin{linenomath}
			\begin{align}
				\label{resolvent_reduced_Z_written}
				\left[(\lambda - \mathcal{L}_{Z})h_{x}\right](s) = g_{x}(s) & & s \in S,
			\end{align}
		\end{linenomath}
		for all $x \in \mathcal{X}$ as $\mathfrak{f}$ is a solution of resolvent equation \eqref{resolvent_reduced_Z}. Since \eqref{resolvent_reduced_Z_written} can be rewritten (cf. \eqref{generatorZ}) as
		\begin{linenomath}
			\begin{align*}
				\left[(\lambda - \mathcal{L})h_{x}\right](s) = g_{x}(s) + (\mathcal{L}_{X}\mathfrak{f})(s,x),
			\end{align*}
		\end{linenomath}
		it follows that $h_{x}(s)$ is also a solution of resolvent equation \eqref{resolvent_reduced_Zmodified} for all $x \in \mathcal{X}$ fixed. By the unicity of the solution, it follows that $f_{x}(s) = h_{x}(s) = \mathfrak{f}(s,x)$ for all $s \in S$ and $x \in \mathcal{X}$, implying \eqref{equality_toShow}.
		
		We turn to the proof of \eqref{limit_toShow}. Observe that, defining $F_{N}(\eta,x) = F_{N,x}(\eta)$ and recalling the definition of $\mathcal{L}_{Z_{N}}$ (cf. \eqref{generatorZN}) and of $\mathcal{L}_{X_{N}}$ (cf. \eqref{generatorXN}),
		\begin{linenomath}
			\begin{align*}
				&\left[(\lambda - \mathcal{L}_{Z_{N}})F_{N}\right](\eta,x) = \left[(\lambda - \mathcal{L}_{N})F_{N,x}\right](\eta) - \left[\mathcal{L}_{X_{N}}F_{N}\right](\eta,x) \\
				&= \left[(\lambda - \mathcal{L}_{N})F_{N,x}\right](\eta) - \sum_{y \in \mathcal{X}} \gamma_{N}(\eta,y) \left[F_{N}(\eta,y) - F_{N}(\eta,x)\right]\\
				&= \left[(\lambda - \mathcal{L}_{N})F_{N,x}\right](\eta) - \sum_{s \in S} \sum_{y \in \mathcal{X}} \gamma(s,y) \left[\mathfrak{f}(s,y) - \mathfrak{f}(s,x)\right] \chi_{\mathscr{E}_{N}^{s}}(\eta) + o_{N}(1)
			\end{align*}
		\end{linenomath}
		in which $o_{N}(1)$ is a quantity that converges to zero when $N$ goes to infinity. Observe that the last equality follows from the fact that $F_{N}(\eta,x) \leq C$ for all $N,\eta,x$ and a constant $C > 0$ (see \cite[(4.2)]{landim2021resolvent}); that $\gamma_{N}(\eta,y)$ converges to $\gamma(s,y)$ for $\eta \in \mathscr{E}_{N}^{s}$ and to zero for $\eta \notin \mathscr{E}_{N}$ (cf. \eqref{limzero} and \eqref{assum_prob}); that $F_{N}(\eta,x)$ converges to $f_{x}(s)$ for $\eta \in \mathscr{E}_{N}^{s}$ (cf. \eqref{implied}); and that $f_{x}(s) = \mathfrak{f}(s,x)$ (cf. \eqref{equality_toShow}). 
		
		By the definition of $\mathcal{L}_{X}$ (cf. \eqref{generatorLX}) it follows that
		\begin{linenomath}
			\begin{align}
				\label{solution_oN1} \nonumber
				\left[(\lambda - \mathcal{L}_{Z_{N}})F_{N}\right](\eta,x) &= \left[(\lambda - \mathcal{L}_{N})F_{N,x}\right](\eta) - \sum_{s \in S} (\mathcal{L}_{X}\mathfrak{f})(s,x) \chi_{\mathscr{E}_{N}^{s}}(\eta) + o_{N}(1) \\ \nonumber 
				&= G_{N,x}(\eta) - \sum_{s \in S} (\mathcal{L}_{X}\mathfrak{f})(s,x) \chi_{\mathscr{E}_{N}^{s}}(\eta) + o_{N}(1)\\
				&= \mathscr{G}_{N}(\eta,x) + o_{N}(1)
			\end{align}
		\end{linenomath}
		in which the second equality follows from the fact that $F_{N,x}$ is the solution of resolvent equation \eqref{resolvent_etaX} and the last equality follows from the definition of $G_{N,x}(\eta)$ and $\mathscr{G}_{N}(\eta,x)$. Subtracting \eqref{resolventZ} from \eqref{solution_oN1} we obtain that, for all $\eta \in \mathcal{H}_{N}, x \in \mathcal{X}$, fixed
		\begin{linenomath}
			\begin{equation*}
				(\lambda - \mathcal{L}_{Z_{N}})\left(F_{N}(\eta,x) - \mathscr{F}_{N}(\eta,x)\right) = o_{N}(1)
			\end{equation*}
		\end{linenomath}
		and \eqref{limit_toShow} follows from the unicity of the resolvent equation solution by making $N \to \infty$.
	\end{proof}
	
	
\end{document}